\documentclass[aps, prb,twocolumn,superscriptaddress]{revtex4-1}

\usepackage{graphicx}
\usepackage{amsmath, bm}
\usepackage{amssymb}
\usepackage{lineno}
\usepackage{ulem}
\usepackage{color}

\newcommand{\pk}{\textbf{k}$_{\mathbf{SAW}}$}
\newcommand{\mkSAW}{$-$\textbf{k}$_{\mathbf{SAW}}$}
\newcommand{\kSAW}{\textbf{k}$_{\mathbf{SAW}}$}

\begin{document}

\title{
Non-symmetric spin-pumping in a multiferroic heterostructure
}

\author{Pauline Rovillain}
\email{pauline.rovillain@insp.upmc.fr}\affiliation{Sorbonne Universit\'e, CNRS, Institut des NanoSciences de Paris, INSP, UMR7588, F-75005 Paris, France}
\author{Ronei Cardoso de Oliveira}
\affiliation{Laboratório de Nanoestruturas para Sensores, Universidade Federal do Paraná, Curitiba 81531-990, PR, Brazil}
\author{Massimiliano Marangolo}
\affiliation{Sorbonne Universit\'e, CNRS, Institut des NanoSciences de Paris, INSP, UMR7588, F-75005 Paris, France}
\author{Jean-Yves Duquesne}
\affiliation{Sorbonne Universit\'e, CNRS, Institut des NanoSciences de Paris, INSP, UMR7588, F-75005 Paris, France}

\date{\today}

\begin{abstract}
We present spin pumping experiments in Co/Pt bilayers grown on Al$_{2}$O$_{3}$(0001) and on ferroelectric Y-cut LiNbO$_{3}$ substrates. Spin pumping is triggered by resonant ferromagnetic resonance induced by surface acoustic waves. We observe that spin pumping efficiency varies when both the magnetization vector and the acoustic wave vector are reversed in  Pt/Co/LiNbO$_{3}$. This phenomenon is not observed in Pt/Co/Al$_{2}$O$_{3}$. We propose that the in-plane electric polarization of LiNbO$_{3}$ is at the origin of the observed phenomenon. These observations open up the perspective of an electric field control of spin pumping efficiency.

\end{abstract}

\maketitle

\section{Introduction}
The first major success in spintronics research was the development of the giant  magneto-resistance spin valve, demonstrating the ability to control a spin-polarized current by taking advantage of the spontaneous imbalance of spin-up and spin-down electrons in the conduction band of ferromagnetic materials \cite{ref:Baibich1988, ref:Binasch1989, ref:George1994}. More recently, pure spin currents, i.e. a spin flux without a corresponding charge flux, are of great interest to the scientific community since they do not give rise to an Oersted field and more importantly, they do not produce Joule heating. Pure-spin current devices are therefore a route to reducing the power consumption of spintronic devices such as, for example, in STT-MRAM\cite{ref:Maekawa2017, ref:Berger1996, ref:Makarov2013}.

The development of low-power spintronic devices based on manipulation of pure spin currents requires an increase in the generation and detection efficiencies via a full understanding of the complex physics behind pure spin currents. Spin current can be detected by the inverse spin Hall effect (ISHE) which corresponds to the conversion of a spin current into a detectable charge current by measuring the voltage on a normal metal (NM) presenting a strong spin-orbit interaction (e.g. Pt). Concerning the generation of pure spin currents different techniques have been proposed in the literature. Emission can be obtained by temperature gradients through the spin Seebeck effect\cite{ref:Uchida2008,ref:Uchida2010,ref:Matsuo2018}, by charge currents (spin Hall effect \cite{ref:Kimura2007}) and by magnetization dynamics (spin pumping \cite{ref:Matsuo2018,ref:Weiler2012, ref:Ando2011}). This last mechanism permits the generation of a pure spin current via ferromagnetic resonance (FMR) in a ferromagnetic material (FM), leading to spin accumulation in an adjacent non magnetic material.

Generally, FMR is obtained by radio-frequency (RF) electromagnetic means or by RF currents \cite{ref:Okada2019}. Interestingly, FMR and spin pumping has also been triggered by acoustic means taking advantage of resonant magnetoelastic coupling in a Co/Pt bilayer\cite{ref:Weiler2012}. Here, mature surface acoustic wave (SAW) technology is used to drive the dynamics of magnetization of a thin Co film, in the GHz regime and by a remote, non inductive and dissipationless method. This so-called SAW-FMR is now well established \cite{ref:Weiler2011, ref:Thevenard2016, ref:Duquesne2019}. The precession pumps a spin current into the NM layer \cite{ref:Adachi2014}. In turn, this spin current generates an electric field ($\mathbf{E}_{\mathrm{ISHE}}$) in the NM layer through the ISHE: $\mathbf{E}_{\mathrm{ISHE}} \sim \mathbf{J}_{S} \times \bm{\sigma}$ where $\mathbf{J}_{S}$ is the spin-current density and $\bm{\sigma}$ correspond to the spin polarization vector. This electric field is converted into a measurable voltage $V_{I}$ between both ends of the Pt strip. What makes SAW assisted experiments extremely interesting is that they are free from spurious non-symmetric signal arising from the anomalous Hall effect (AHE) which affects the standard FMR technique \cite{ref:Saitoh2006}.

The main purpose of this letter is to compare the spin pumping excitation/detection efficiency when both directions of the magnetization {\bf M} and acoustic wave-vector \kSAW~are reversed. We investigated SAW-FMR assisted spin pumping in a multiferroic system where a FM/NM (Co/Pt) bilayer is in contact with a ferroelectric (FE) LiNbO$_{3}$ substrate. We put forward the hypothesis that the remanent electrical polarization vector affects the ISHE voltage.

\section{Samples characteristics and experimental setup} 
\begin{figure}[h!]
\begin{center}
	\includegraphics[width=7.5cm]{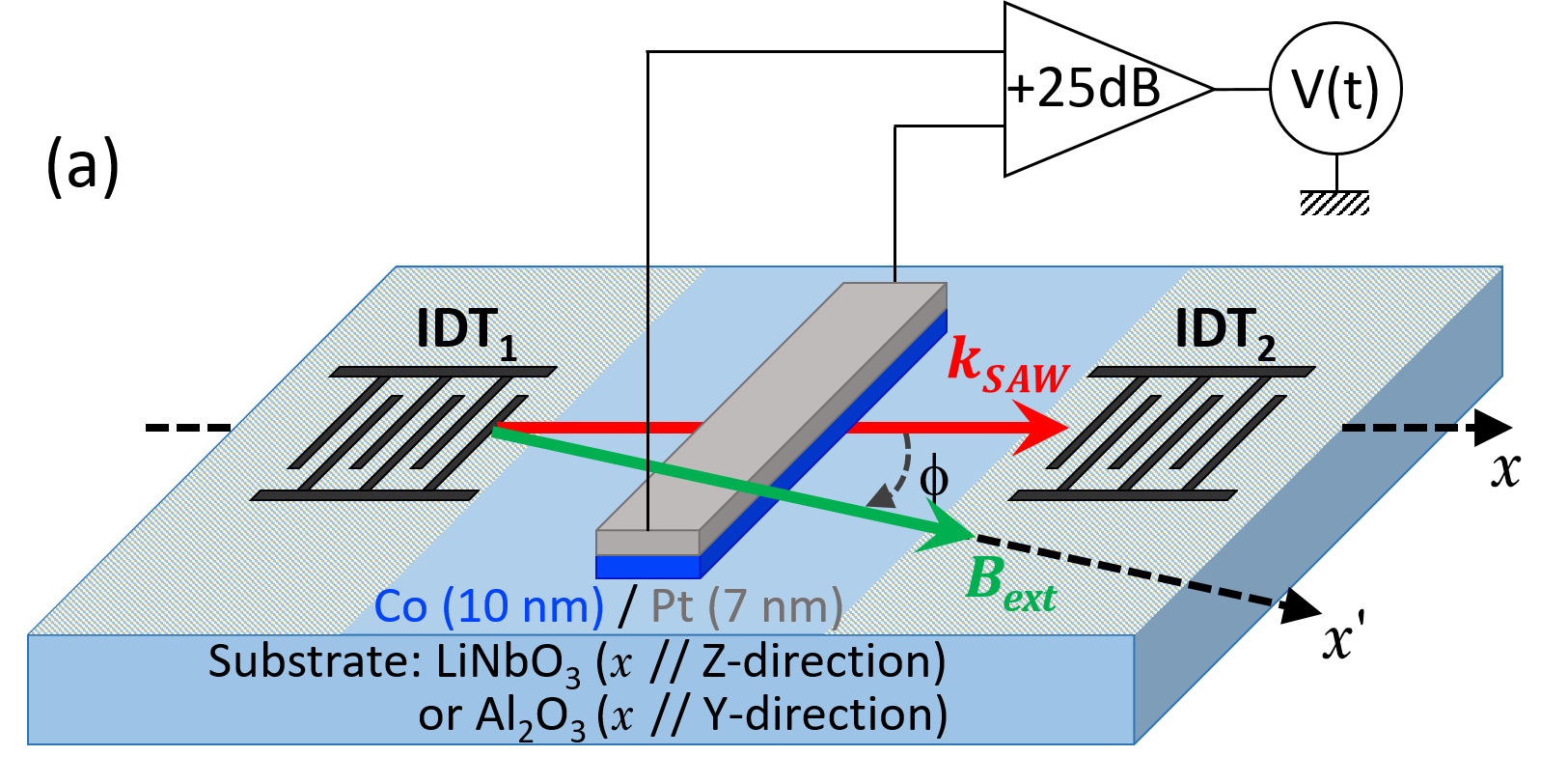}
	\includegraphics[width=6cm]{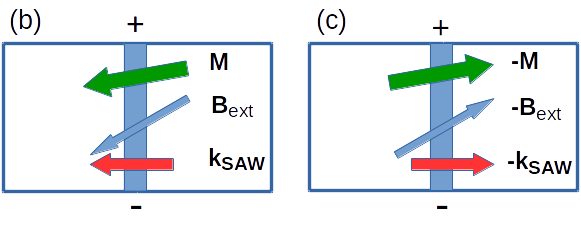}
	\includegraphics[width=8.5cm]{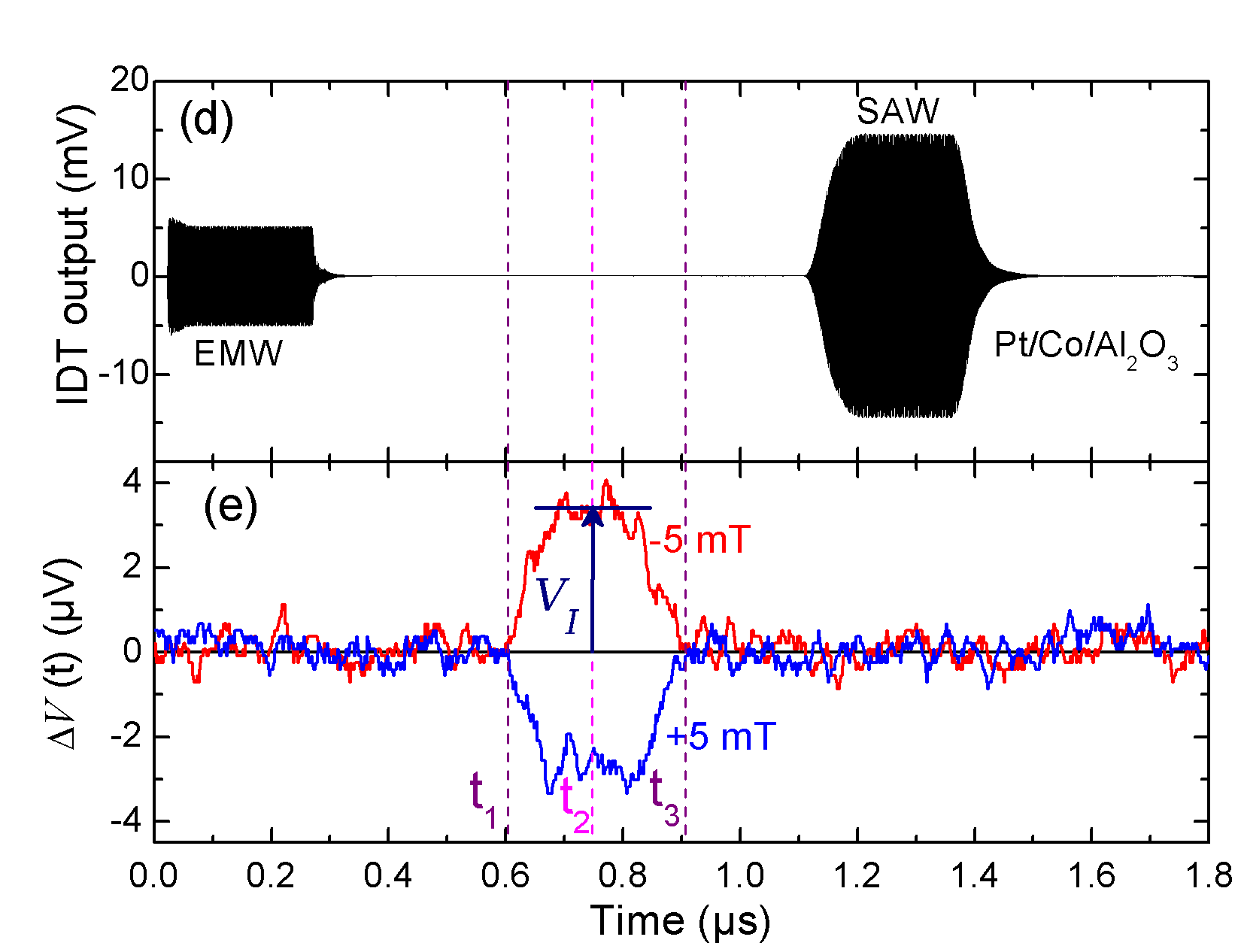}
	\caption
	{
	(a) Sketch of the sample and setup. On the LiNbO$_{3}$ and Al$_{2}$O$_{3}$ samples, IDTs have 30 pairs of fingers with an aperture of 500~$\mu$m. They are made of Al with a metallization ratio equal to 0.5. 
	On the LiNbO$_{3}$ sample, the distance between IDTs is 3~mm and between one IDT and the Co/Pt bilayer is 1.25~mm. On the Al$_{2}$O$_{3}$ sample, the distance between IDTs is 6~mm and between one IDT and the Co/Pt bilayer is 2.75~mm. 
	(b,c) display the sample, with two configurations of in-plane external magnetic field {\bf B$_\mathbf{ext}$}, magnetization {\bf M} and \kSAW.
	(c) is obtained from (b) by inversion of {\bf B$_\mathbf{ext}$}, {\bf M} and \kSAW.
    Spin current signal $V$ is measured between the (+) and (-) terminals.
	(d) Output of the receiving IDT as a function of time. EMW signal and SAW pulses are observed.
	(e) $\Delta V(t,B_{ext}) = V(t,B_{ext}) - V(t,B_{ref}$) where $V$ is the spin current signal. $B_{\mathrm{ext}} = \pm 5$ mT. $B_{\mathrm{ref}} = \pm 35$ mT (see main text).
	Pt/Co/Al$_{2}$O$_{3}$ sample. 
	}
	\label{fig:manip}
\end{center}
\end{figure}

Figure \ref{fig:manip}(a) is a sketch of the samples and setup. We used two different substrates: Y-cut LiNbO$_{3}$ and (0001)-cut Al$_{2}$O$_{3}$. Moreover,  LiNbO$_{3}$ is not centrosymmetric (space group is $R3m$). It is ferroelectric and therefore piezoelectric. We verified that our LiNbO$_{3}$ substrate is a single-domain ferroelectric using polarized light microscopy and piezoelectric force microscopy (PFM) measurements (see appendix \ref{ap:polar}, Fig. \ref{polar2} and \ref{PFM}). No domain wall was observed from large scale (2.86 mm $\times$ 2.11 mm) to low scale (1 $\mu$m $\times$ 1 $\mu$m) that is expected in a single-domain ferroelectric material.  Al$_{2}$O$_{3}$ is centrosymmetric (space group is $R\overline{3}m$). It is neither piezoelectric nor ferroelectric. The orientations of both substrates have been verified by XRD. Both samples do not present a 2-fold rotational symmetry about the surface normal. On both substrates, a $500\mu\text{m}\times1.5\text{mm}$ bi-layer of Co ($10\text{nm}$) and Pt ($7\text{nm}$) is deposited by electron beam evaporation at the center. SAW bursts are emitted and detected electrically by means of two nominally identical IDTs defined by electron beam lithography and placed symmetrically on either side of the central Co/Pt strip. In the case of the Pt/Co/LiNbO$_{3}$ sample, IDTs are deposited directly onto the piezoelectric substrate. However, in case of the Pt/Co/Al$_{2}$O$_{3}$ sample, a thin ZnO piezoelectric layer of 490~nm is deposited on the surface by sputtering at 300$^\circ$C only under the IDTs areas so as to emit SAWs. ZnO is a textured polycrystal with c-axis normal to the substrate.
200ns SAW bursts, with a repetition rate of 5 kHz, are emitted at 1.5 and 1.3 GHz for LiNbO$_{3}$ and Al$_{2}$O$_{3}$ respectively. The electrical power applied during the SAW bursts was 19~dBm (79 mW), except for power study in Fig~\ref{fig:V_vs_P}. After propagation in the Co/Pt layers the SAW is detected by the opposite IDT and the acoustic signal is acquired with a digital oscilloscope. Simultaneously, the output voltage $V(t)$ (between the both ends of the Co/Pt bi-layer) is  measured  with a DC-amplifier (bandwidth = 0-20 MHz, gain = 25 dB, input and output impedance = 50~$\Omega$) and recorded with the same oscilloscope. To improve the signal-to-noise ratio, a sequence of 64000 $V(t)$ signals is averaged.

On the Y-cut LiNbO$_{3}$ substrate, the SAW wave-vector \kSAW~is along the Z-axis ($[0001]$ direction), i.e. the spontaneous direction of the electrical polarization vector $\mathbf{P}$ \cite{ref:Nassau1966a, ref:Nassau1966b}. On the Al$_{2}$O$_{3}$ substrate, \kSAW~is along the Y-axis ($[1\bar{1}00]$ direction). On both substrates we define $x$ parallel to \kSAW, oriented from IDT$_{1}$ to IDT$_{2}$. We also define an in-plane axis $x'$ at an angle $\phi$ from $x$. We apply a magnetic field $\mathbf{B_{ext}}$ along $x'$ (see Fig.\ref{fig:manip}(a)).

The DC spin current is detected via the ISHE in the Pt layer by measuring the time-dependent voltage between both ends of the Co/Pt bi-layer. The output voltage $V(t)$ is recorded synchronously with the acoustic bursts (see Fig.\ref{fig:manip}(d,e)). To separate, in the time domain, the spin current signal $V$ from the electromagnetic wave (EMW) and from the acoustic (SAW) signals, the distances between each IDT and the bi-layer is carefully designed (1.25 and 2.75 mm for LiNbO$_{3}$ and Al$_{2}$O$_{3}$, respectively). In an experimental run, we first set the orientation $\phi$ and we saturate the magnetic film with a large field amplitude $B_\mathrm{sat}$. Then, the field either increases (from -35 to +35 mT, $B_\mathrm{sat}$ = -350~mT) or decreases (from +35 to -35 mT, $B_\mathrm{sat}$ = +350~mT). $V(t)$ is recorded at each field value.
These measurements were performed for a range of orientations $\phi$ and also for opposite acoustic wave-vectors \kSAW. The direction of \textbf{k}$_{\mathbf{SAW}} $ is reversed by connecting the input signal to one IDT or to the other. Since the IDTs are symmetrical with respect to the Co/Pt strip, spin pumping occurs in the same time range $[t_{1},t_{3}]$ for both propagation directions.

For each sample measured, the Co layer was first characterised since each evaporated layer has slightly different characteristics (see appendix \ref{app:magprop}). In this paper, we choose to present the results obtained for two specific samples where the coercive field $B_{c}$ is lower than the resonant field $B_{res}$.

From broadband-FMR (BB-FMR) measurements between 3 and 8.5~GHz, we record the FMR frequency, $F_\mathrm{FMR}$, as a function of the in-plane magnetic field amplitude ($B_{ext}$) \cite{ref:Hamida2013}. By extrapolation of our experimental values ($F_\mathrm{FMR}$ vs $B_{ext}$) using the Kittel formula \cite{ref:Kittel1951}, we found a SAW-FMR resonance condition around 4 mT and 6 mT for Pt/Co on LiNbO$_{3}$ and Al$_{2}$O$_{3}$, respectively. Both samples are invariant by a rotation of $\pi$, regarding their MOKE magnetic cycles (see appendix B and insert of Fig.\ref{kerr}). First, we present our results. Afterward, we will analyse them in term of symmetries.

\section {Results and discussion}

Measurements, presented here, have been performed at $\phi = 0^\circ$ with $B_{c}$ = 1.1~mT for Pt/Co/LiNbO$_{3}$ and $\phi = 60^\circ$ with $B_{c}$ = 3~mT for Pt/Co/Al$_{2}$O$_{3}$.
Figure \ref{fig:manip}(e) shows the spin current signal $\Delta V(t) = V(t,B_{ext})-V(t,B_{ref})$  for Pt/Co/Al$_{2}$O$_{3}$, $B_{\mathrm{ext}} = \pm 5$ mT and $B_{\mathrm{ref}} = \pm 35$ mT. This subtraction is performed to eliminate parasitic deterministic signals.
According to calculation, no SAW-FMR and no spin pumping are expected at $\pm 35~\mathrm{mT}$.  On the contrary, at $B_{\mathrm{ext}} = \pm 5$ mT, the acoustic frequency matches the precession frequency of the uniform mode (See appendix B, Fig.~\ref{bbfmr}). Hence, when the acoustic bursts go through the Co/Pt bi-layer, in the time range $[t_{1}$,$t_{3}]$ ([0.6-0.9] $\mu$s in the case of the Pt/Co/Al$_{2}$O$_{3}$ sample, Fig. \ref{fig:manip} (e)), they induce the ferromagnetic resonance of Co and a pure spin current is launched in the adjacent Pt layer, with a maximum value around $t_{2}$. The magnitude of the spin current is proportional to the time-averaged value $V_{I}$ of $\Delta V(t)$, around $t_{2}$ (between 0.7 and 0.8~$\mu$s for Pt/Co/Al$_{2}$O$_{3}$). 

\begin{figure}
	\includegraphics[width=\linewidth]{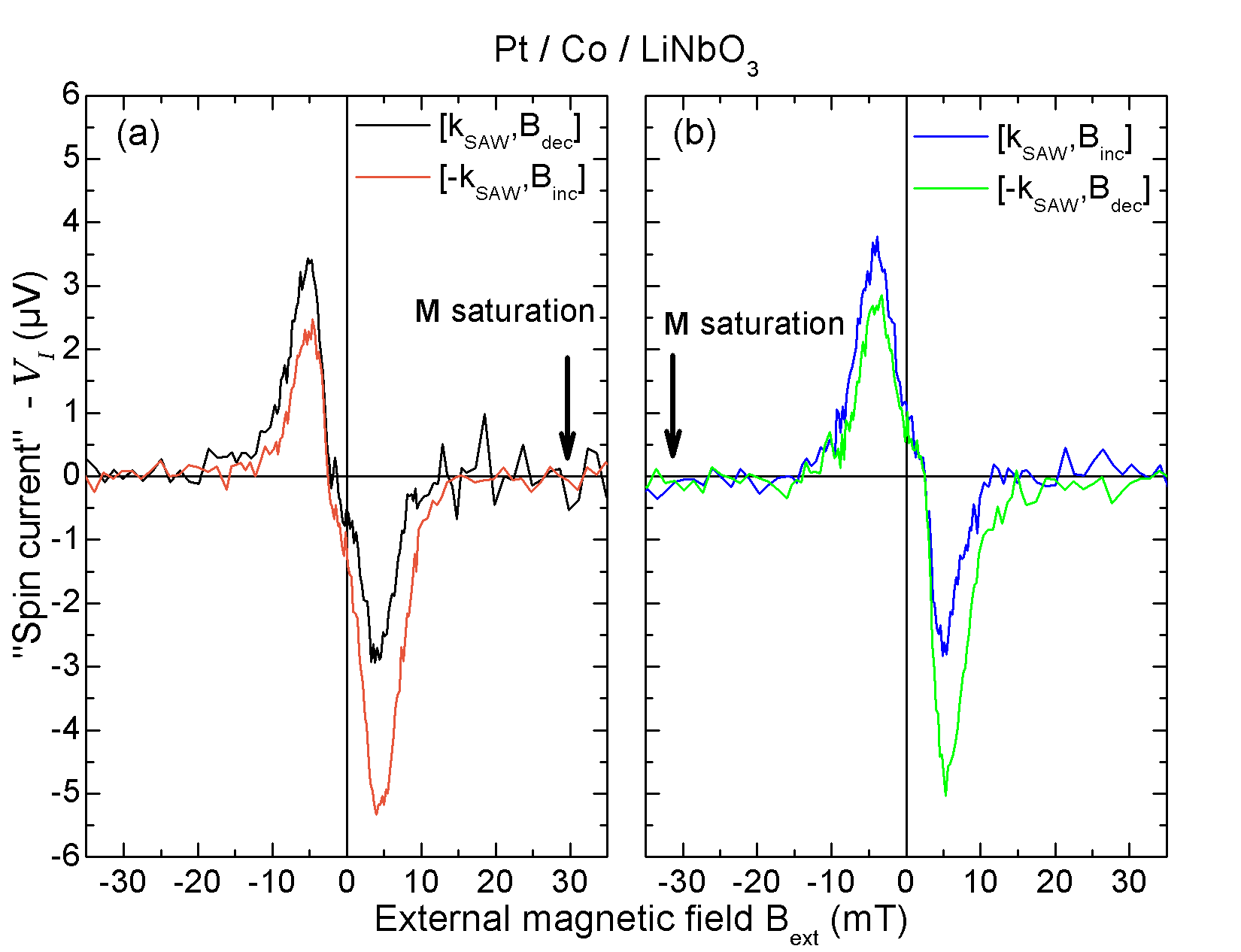}
	\includegraphics[width=\linewidth]{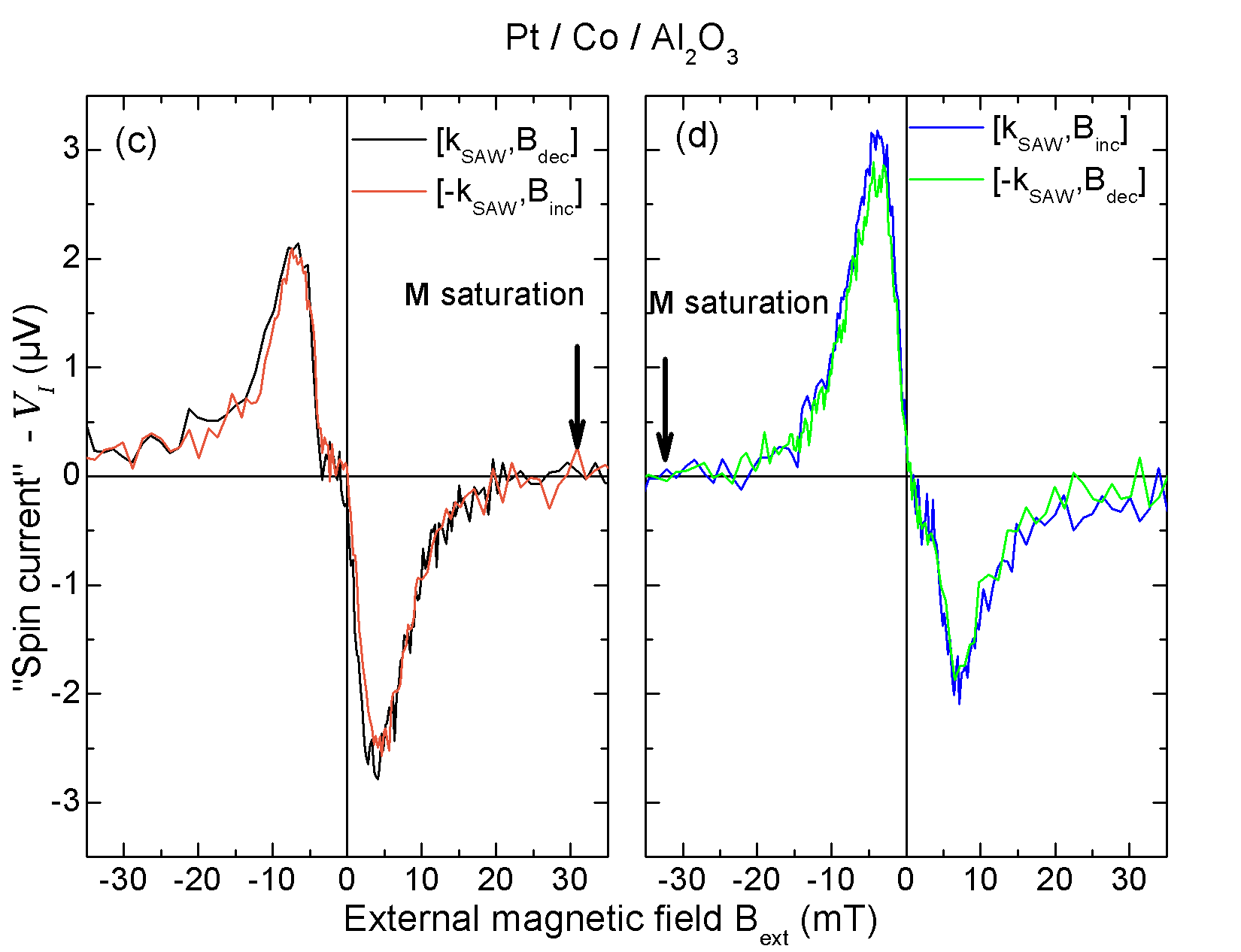}
	\caption{
	(a, b) Co/Pt on LiNbO$_{3}$ and (c, d) Co/Pt on Al$_{2}$O$_{3}$.
	$V_{I}$ (proportional to the spin current) versus $\mathrm{B_{ext}}$ after symmetry operations on \mkSAW~scans. Black and blue curves: raw data. 
	Orange and green curves: starting from raw data, $V_{I}$ and $B_{ext}$ are multiplied by -1 (see main text).
	(a, c) Comparison between [\pk, $B_\mathrm{dec}$] and [\mkSAW, $B_\mathrm{inc}$].
	(b, d) Comparison between [\pk, $B_\mathrm{inc}$] and [\mkSAW, $B_\mathrm{dec}$].
	}
	\label{fig:AlO_invers} 
\end{figure}

We assume that the spin current is maximum when the resonance is triggered. Then, we identify the resonance field $B_{res}$ at the value where the current signal $V_{I}$ is maximum (see Fig. \ref{fig:AlO_invers}). We observe the change of sign of $V_{I}$ with the reversal of $\mathbf{M}$, induced by the reversal of $\bf B_{ext}$ (since $B_{c} < B_{res}$), which is an expected condition for acoustic spin pumping \cite{ref:Czeschka2011}.
\begin{figure}
	\includegraphics[width=\linewidth]{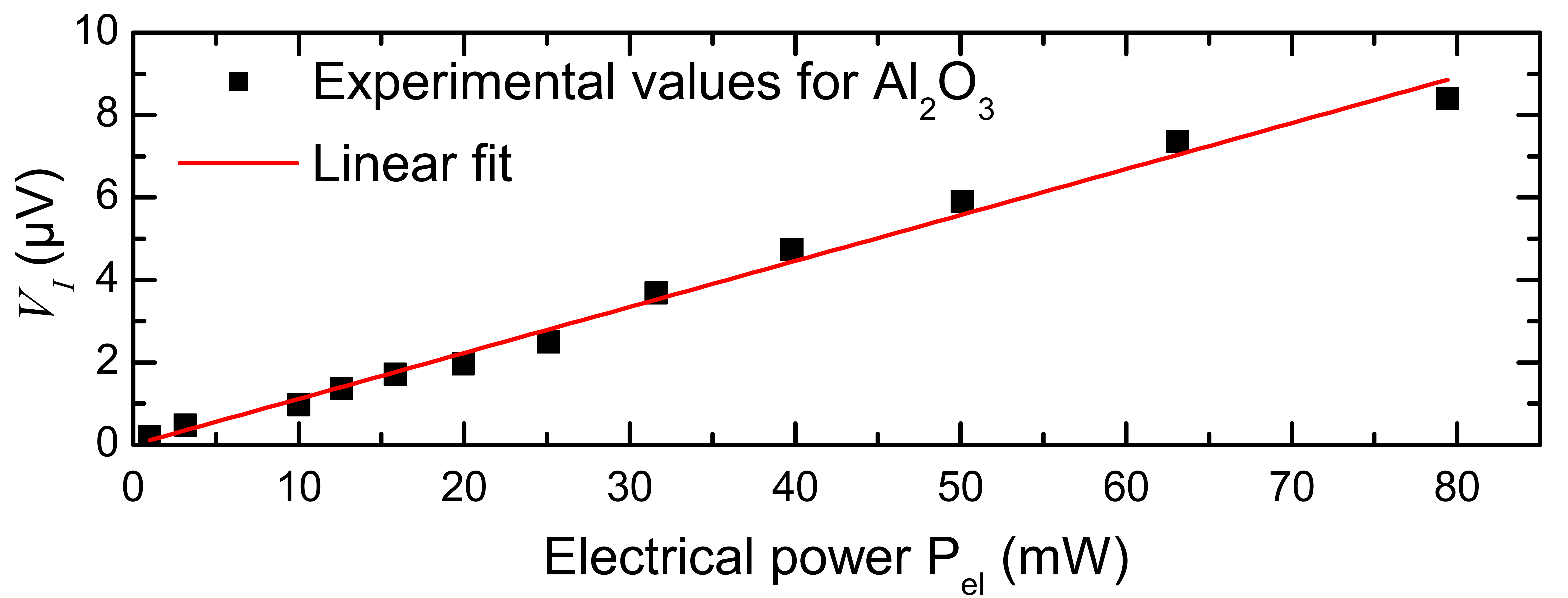}
	\caption
	{
	$V_{I}$ (proportional to the spin current amplitude) as function of incident electrical power on IDT for Pt/Co/Al$_{2}$O$_{3}$ sample  \cite{ref:footnote1}. The red line is a linear fit.
	}
	\label{fig:V_vs_P}
\end{figure}
We also note that $V_{I}$ (proportional to the spin current) is proportional to the incident electrical power ($P_{el}$), and then to the electrical power injected in the IDT, as shown in Fig.\ref{fig:V_vs_P} for the bilayer on the Al$_{2}$O$_{3}$  substrate. The same behavior is observed for the LiNbO$_{3}$ sample. Indeed, this is in good agreement with experiments of K.Ando {\it et al.}\cite{ref:Ando2011} and with the model of F.Czeschka {\it et al.}\cite{ref:Czeschka2011}, from which we can easily derive, in the small angle regime: $V_I \propto P_{el}$.

At first glance (see Fig.\ref{fig:AlO_invers}), the $V_{I}$ dependence on {\bf B}$_\mathrm{ext}$ shown here seems to be similar to that shown by M.Weiler {\it et al}\cite{ref:Weiler2012}. However, a difference is observed in our measurements at the coercive fields. Indeed, a significant electric current is emitted at $-B_{c}$ and $+B_{c}$ for decreasing and increasing field respectively, which is difficult to observe in M.Weiler {\it et al}\cite{ref:Weiler2012}. The origin of this current is not yet understood and is beyond the scope of this paper. However, since $B_{c}$ is close to the resonant field, the current superimposes on the spin current and it becomes difficult to separate the two contributions. To avoid the spurious contributions, we concentrate only on the measurements of $V_{I}$ for low coercive field samples and at angles $\phi$ where $B_{c}$ is weakest.

So, let us now see what is expected when the magnetic field and the acoustic wave-vector are reversed. Figures \ref{fig:manip}(b,c) displays sketches of a sample (substrate and bi-layer) subject to two configurations (b) and (c), i.e. different orientations of in-plane applied field \textbf{B}$_{\mathbf{ext}}$, magnetization \textbf{M} and acoustic wave-vector \kSAW. Configuration (c) is obtained from (b) by reversing \textbf{B}$_{\mathbf{ext}}$, \textbf{M} and \kSAW. A trivial behaviour is expected for systems that are invariant by a rotation of $\pi$ about an axis perpendicular to the propagation plane. In that case, we can notice that configuration (c) can also be obtained from (b) by combining a rotation of $\pi$ and an exchange of the (+) and (-) terminals. Therefore, if the sample is invariant by a rotation of $\pi$, we infer that going from (b) to (c) will change the sign of the measured voltage between the (+) and (-) terminals but not its magnitude. 


From an experimental point of view, reversing \kSAW~ and \textbf{B}$_{\mathbf{ext}}$ is quite easy. However, reversing \textbf{M} is a little bit more tricky because of the memory effect associated with the magnetic hysteresis. In Fig.\ref{fig:manip}, if configuration (b) is reached from a positive saturating field ("decreasing field scan"), configuration (c) must be reached from a negative saturating field ("increasing field scan"), and conversely. To conclude, in order to check the symmetry rules, we should compare the voltages measured either for [\kSAW, \textbf{B}$_{\mathbf{ext}}$, $B_\mathrm{dec}$] and [\mkSAW, $-$\textbf{B}$_{\mathbf{ext}}$, $B_\mathrm{inc}$], or for [\kSAW, \textbf{B}$_{\mathbf{ext}}$, $B_\mathrm{inc}$] and [\mkSAW, $-$\textbf{B}$_{\mathbf{ext}}$, $B_\mathrm{dec}$]. From a practical point of view, the comparison can be easily carried out in the following manner. Let us consider two field scans, [\kSAW, $B_\mathrm{inc}$] and  [\mkSAW, $B_\mathrm{dec}$]. For the first scan, we plot the voltage $V_{I}$ versus the signed  amplitude of the field. For the second scan, we plot the opposite voltage $-V_{I}$ versus the opposite of the signed amplitude of the field.

Fig.\ref{fig:AlO_invers} (a,b) displays our results for the Pt/Co/LiNbO$_{3}$ sample. Clearly, the curves obtained for [\kSAW, $B_\mathrm{dec}$] and  [\mkSAW, $B_\mathrm{inc}$] scans (or [\kSAW, $B_\mathrm{inc}$] and  [\mkSAW, $B_\mathrm{dec}$] scans) do not superimpose. For example, at $B_{ext} = 5$~mT, the magnitude of the voltage is either 2.5 or 5 $\mu$V, depending on the scans. This factor of 2 cannot be attributed to possible different intensities of the acoustic field, at \pk~and \mkSAW, arising from different efficiencies of the IDT$_{1}$ and IDT$_{2}$ transducers. If a multiplying factor were applied to correct the amplitudes at 5~mT, then the difference in amplitudes at -5~mT would further increase. 
We observed this for all angles $\phi$ and samples studied even if $B_{c} > B_\mathrm{res}$. This phenomenon must not be confused with "non-reciprocal" acoustic propagation observed in magnetic materials where only \pk\, is inverted \cite{ref:Camley1987}.

Fig.\ref{fig:AlO_invers} (c,d) displays our results for the Pt/Co/Al$_2$O$_{3}$ sample. The curves obtained for corresponding field scans superimpose. Reversing simultaneously \kSAW, \textbf{B}$_{\mathbf{ext}}$ and \textbf{M} changes the sign of the measured voltage but not its magnitude. Indeed, slightly different amplitudes are observed which can be attributed to different intensities of the acoustic field, at \pk~and \mkSAW, arising from different efficiencies of the two transducers. The same multiplying factor of 1.1 can be applied to correct the amplitudes at 5~mT and at -5~mT. 

In the Pt/Co/LiNbO$_{3}$ sample, the origin of the effect has to be determined. Fundamentally, the non-symmetry of the current originates from the non-invariance of the system by a rotation of $\pi$ (about an axis perpendicular to the propagation plane). A tentative explanation involves the polarization vector, $\mathbf{P}$, that is oriented along the in-plane Z direction. One possibility concerns the excitation process and involves a coupling between the ferroelectric polarization $\mathbf{P}$ of LiNbO$_{3}$ and the ferromagnetic moment of Co, as seen in magnetoelectric materials \cite{ref:Velev2011} or in artificial composite multiferroic systems \cite{ref:Jedrecy2013, ref:Jia2014, ref:Chiba2016} where the $\mathbf{P}$ vector is oriented perpendicularly to the FE/FM interface, permitting an efficient coupling of ferroic orders. We point out that in Y-cut LiNbO$_{3}$ substrates, $\mathbf{P}$ lies parallel to the interface. In that configuration, the electrical field in the cobalt thin film goes to zero. Nevertheless, we put forward the hypothesis that the interface roughness can induce interface charge accumulation and consequent non equilibrium spin density along the out-of-plane direction. As pointed out in Jia {\it et al.}\cite{ref:Jia2014}, this spin-unbalanced interface electrostatic effect accompanied by the so-called s-d exchange interaction, may give rise to an effective magneto-electric (ME) coupling affecting the whole Co thin film and the consequent spin pumping efficiency at the Co/Pt interface. This could modify either the magnetic precession cone angle at resonance, or the magnetic precession frequency, or both, and then modify the spin current generation. Nevertheless the BB-FMR measurements that we performed did not show any evidence of an FMR resonance shift after magnetic field reversal. This hypothesis can be tested in the future by a systematic study of non-symmetric spin pumping as a function of the interface roughness and/or the misalignment of the normal to the surface. 
To compare with the experiments performed on Pt/Co/Al$_{2}$O$_{3}$ (which display invariance by a $\pi$-rotation), we notice that the Al$_{2}$O$_{3}$ structure is also not invariant by a $\pi$-rotation since the symmetry about (0001) is only of order 3. However, a $2\pi/6$ rotation leaves unchanged the location of the oxygen atoms and only involves a change in the orientation of the aluminium triangles which are located between the oxygen planes. Therefore, in our Pt/Co/Al$_{2}$O$_{3}$ sample, the non-invariance by a rotation of $\pi$ is somehow "weak" since it is just due to slight variations of atoms stacking.

Other processes involving the detection can be evoked due to the electrical field in the Pt layer, arising from $\mathbf{P}$. However, they must be very weak, because of the geometry and of screening effects.
In order to corroborate our hypothesis it would be interesting to quantify the effect of electrical polarization on spin current generation by modifying or reversing {\bf P} through the application of an electric field. Unfortunately, the fields required to reverse the polarization are beyond our technical capabilities. Indeed, the coercive electric field of the LiNbO$_{3}$ is $E_{c} > 210$ kV.cm$^{-1}$ \cite{ref:Kim2002} which would require to apply on our samples a voltage $U > 21$ kV.

\section{Conclusion}
In conclusion, we have investigated the acoustic spin pumping in Co/Pt bilayers deposited on LiNbO$_{3}$ and Al$_{2}$O$_{3}$ substrates. Our experiments pinpoint the lack of $\pi$ rotational symmetry of spin-pumping signal that we possibly attribute to the permanent electrical polarization of LiNbO$_{3}$. We suggest that this  polarization is an important factor in the spin pumping excitation processes. The nature of the coupling with the magnetic polarization is still unclear. It may be due to a magnetoelectric coupling at the interface between LiNbO$_{3}$ and Co, inducing non equilibrium spin density across the whole thin film \cite{ref:Jia2014} and affecting spin pumping efficiency.
It will be interesting to use a ferroelectric substrate whose polarisation {\bf P} can be easily varied, in order to modify the spin current emission in a bilayer.

\begin{acknowledgments}
The authors acknowledge L.Becerra and M.Rosticher for optical and electronic lithography, S.Suffit for e-beam evaporation of Co/Pt bilayers, F.Vidal for assistance during MOKE experiments, Y.J. Zheng for X-ray diffraction measurements, H.Cruguel for PFM measurements and C.Gourdon, L.Thevenard and P.Atkinson for careful reading.
\end{acknowledgments}

\appendix

\section{Single domain evidence in ferroelectric LiNbO$_3$}
\label{ap:polar}

\begin{figure}
	\includegraphics[width=\linewidth]{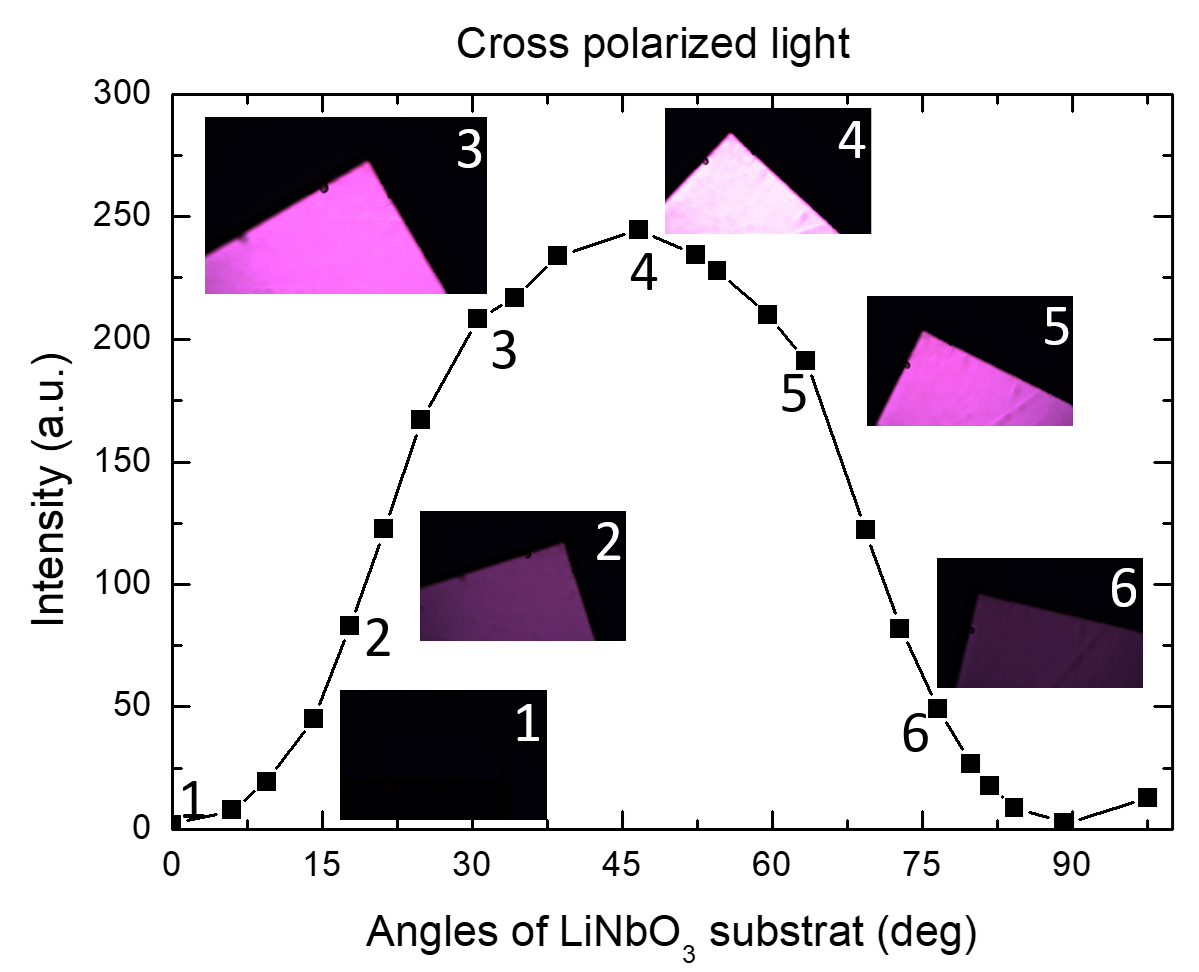}
	\caption{Cross-polarized light intensity transmitted through a LiNbO$_{3}$ substrate, as a function of the angle of the substrate. Inset images: pictures under cross polarized light microscopy (2.86 mm $\times$ 2.11 mm). Numbers link images and intensities, at a given angle.
	}
	\label{polar2}
\end{figure}

\begin{figure}
	\includegraphics[width=\linewidth]{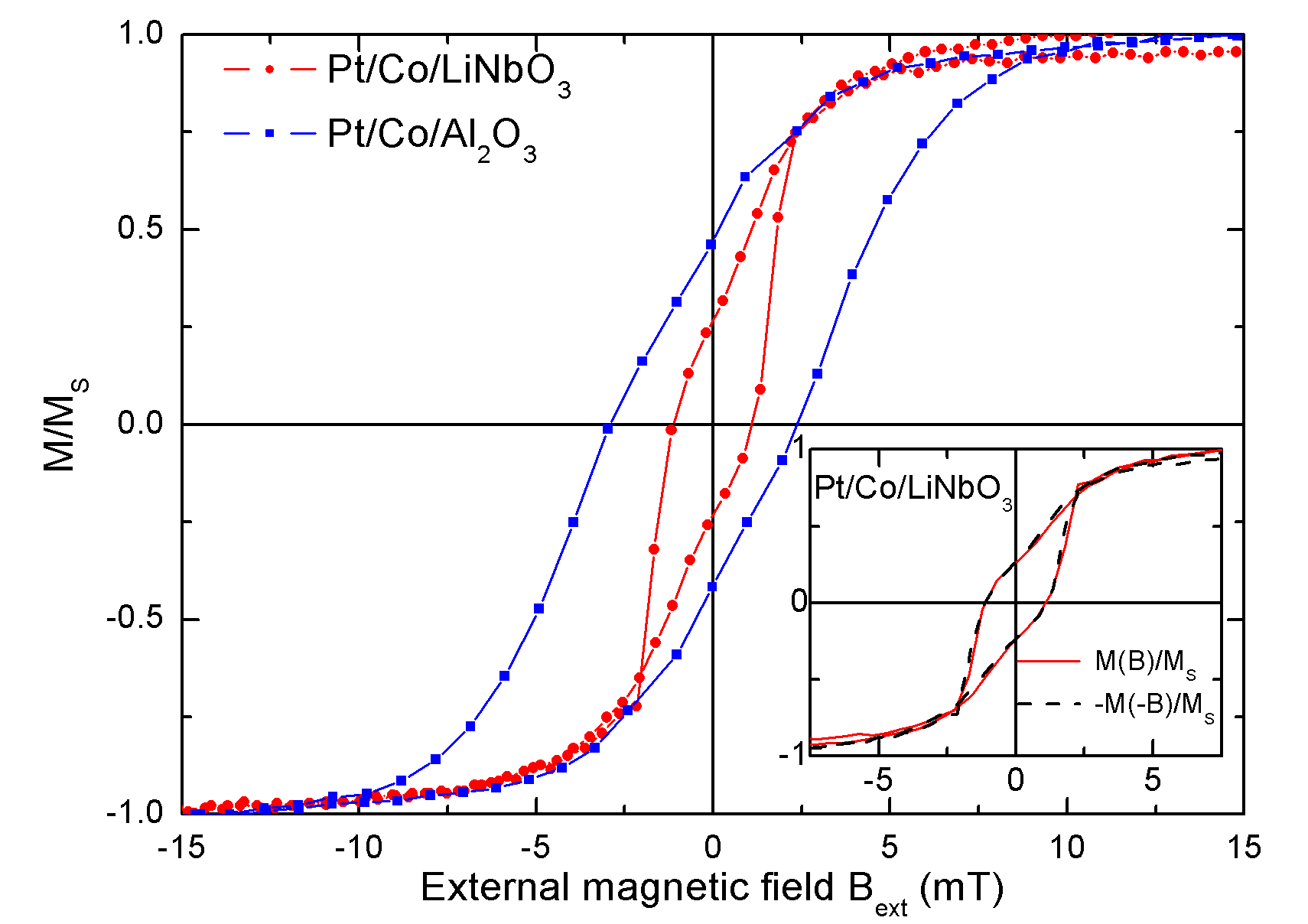}
	\caption{PFM images of a surface of LiNbO$_{3}$ substrate. (a,b,c): 20 $\mu$m $\times$ 20 $\mu$m image size, (d,e,f): 1 $\mu$m $\times$ 1 $\mu$m image size. (a,d) correspond to topography measurements of the surface. (b,e) show the response of out of plane piezoelectricity and (c,f) the in plane piezoelectricity one.
	}
	\label{PFM}
\end{figure}

To verify that the LiNbO$_{3}$ substrate is a single ferroelectric domain, we observed a substrate (from a series) using polarized light microscopy. Fig.\ref{polar2} summarizes our observation on a large scale (2.86 mm $\times$ 2.11 mm) for a range of angles, in a rotation around the normal to the substrate. As expected, the intensity of the transmitted light depends on the angle. However, no contrast is observed over the sample, suggesting a single domain configuration. Nevertheless, anti-parallel domains cannot be excluded. To clear this point, we observed the sample (using polarized light microscopy) on a much smaller scale (110 $\mu$m $\times$ 83 $\mu$m) in an attempt to reveal walls between anti-parallel domains. According to V.G.Zalessky {\it et al.}\cite{ref:Zalessky2006}, domain walls induce optical contrast on this scale.  Moreover, piezoelectric force microscopy (PFM) measurement is performed in order to detect micro- and/or nano-metric ferroelectric domains. Figures \ref{PFM} (a,b,c) show PFM images for a large scale (20 $\mu$m $\times$ 20 $\mu$m) and (d,e,f) show PFM images for a smallest scale (1 $\mu$m $\times$ 1 $\mu$m). It turns out that ou-of-plane and in-plane measurements do not put into evidence any ferroelectric domain for both scale. We observe only a line which is present as well in the topographic and piezoelectric mode indicating that this line is a defect on the surface. Since we observed neither walls and domains, we conclude that our LiNbO$_{3}$ substrates exhibit a single ferroelectric domain configuration\cite{ref:Zalessky2006, Kiselev2012}.

\section{Magnetic properties}
\label{app:magprop}
\begin{figure}
	\includegraphics[width=\linewidth]{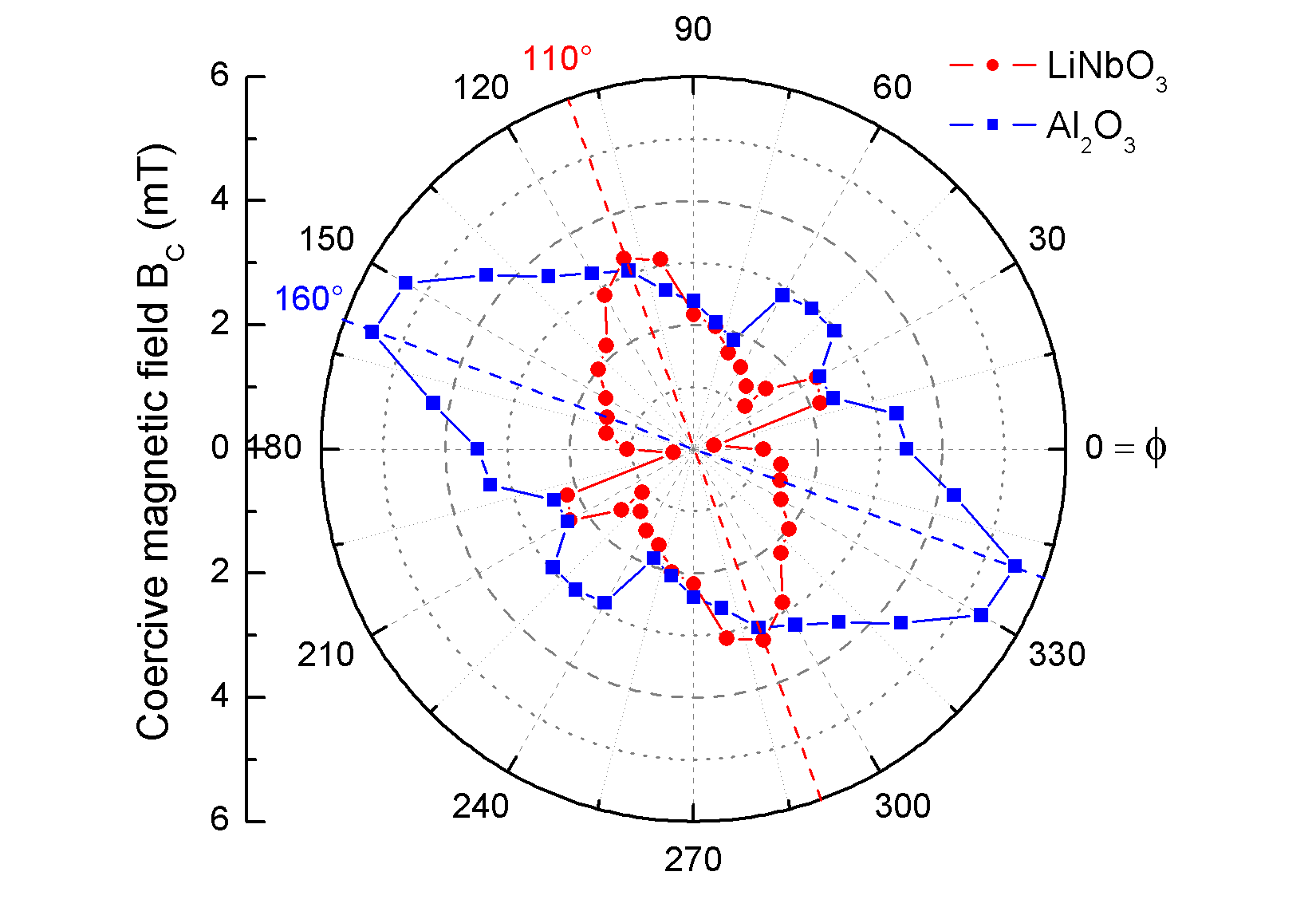}
	\caption{
		Magnetic cycles $M(B)/M_{S}$ for Pt/Co/LiNbO$_{3}$ (at $\phi = 0^\circ$) and Pt/Co/Al$_{2}$O$_{3}$~(at $\phi = 60^\circ$) samples.
		Insert: Pt/Co/LiNbO$_{3}$ (at $\phi = 0^\circ$). Comparison of $M(B)/M_{S}$ and $-M(-B)/M_{S}$.
	}
	\label{kerr}
\end{figure}

From the magnetic point of view, each evaporated Co layer has different characteristics. Figure \ref{kerr} shows the magnetic cycles $M(B)/M_{S}$ for Co layers deposited on LiNbO$_{3}$~(at $\phi = 0^\circ$) and Al$_{2}$O$_{3}$~(at $\phi = 60^\circ$) substrates mesured by magneto-optical Kerr effect. From this measurement we can extracted the coercive field: $B_{c}$ = 1.1~mT on Pt/Co/LiNbO$_{3}$ \, and $B_{c}$ = 3~mT on Pt/Co/Al$_{2}$O$_{3}$.
The insert of Fig.\ref{kerr} displays the raw and processed data, in case of LiNbO$_{3}$: $M(B)/M_{S}$ and $-M(-B)/M_{S}$, respectively. The nearly perfect superposition indicates that the magnetic layer is invariant by a rotation of $\pi$ (same for Co on Al$_{2}$O$_{3}$, not shown).\\

\begin{figure}
	\includegraphics[width=\linewidth]{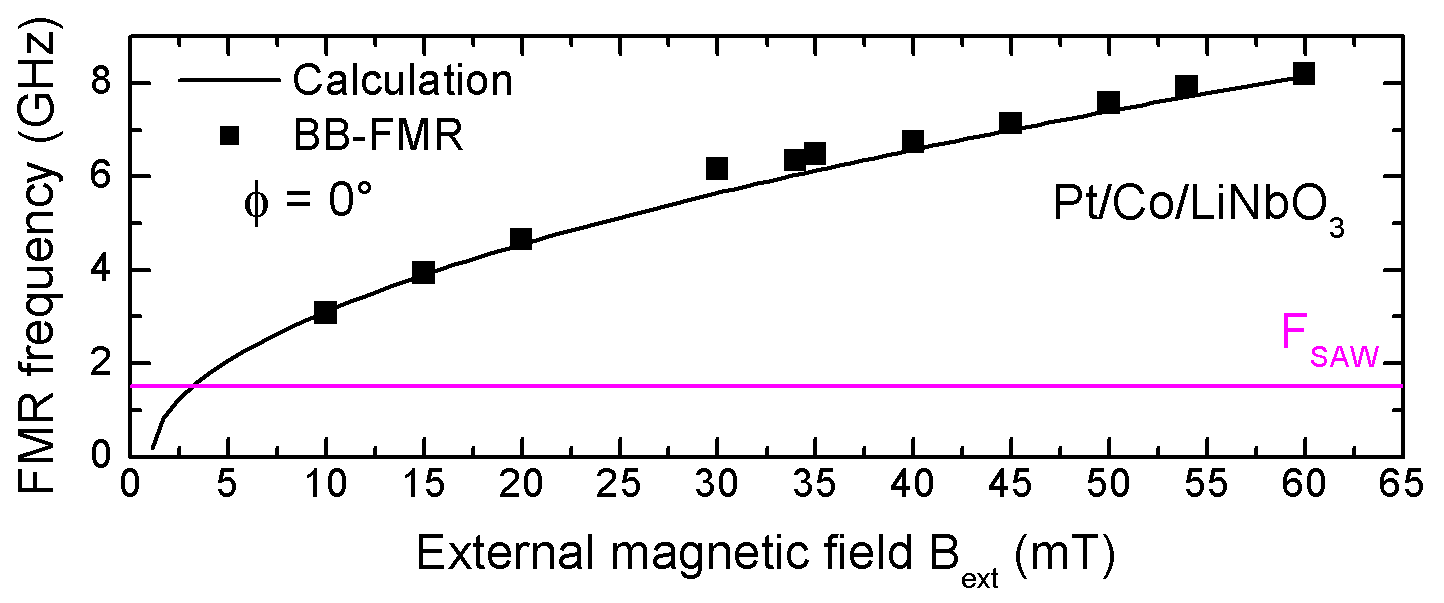}
	\caption{
		Coercive magnetic field versus angle $\phi$ between the magnetic field and \pk~direction for the Co layer on Al$_{2}$O$_{3}$~(blue squares) and LiNbO$_{3}$~(red dots).
	}
	\label{moke}
\end{figure}

The magnetic cycle is recorded for different angles. Figure \ref{moke} displays $B_{c}$ as function of the angle, we can see that the magnetic properties are different between the two Co layers as the easy axis and the magnitude of $B_{c}$ do not match. \\

\begin{figure}
	\centering
	\includegraphics[width=\linewidth]{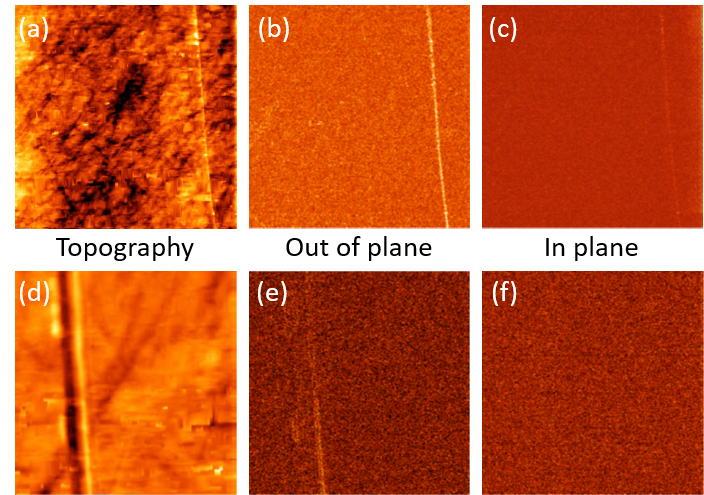}
	\caption{(black square) Experimental FMR frequencies of the Co layer on LiNbO$_{3}$~as function of the external magnetic field. (black curve) Extrapolation by the Kittel formula. (Magenta curve) SAW excitation frequency.}
	\label{bbfmr}
\end{figure}

Figure \ref{bbfmr} represents the FMR frequency, $F_\mathrm{FMR}$, as a function of the in-plane magnetic field amplitude ($B_\mathrm{ext}$), from broadband-FMR (BB-FMR) measurements between 2.5 and 8.5~GHz. By extrapolation of our experimental values ($F_\mathrm{FMR}$ vs $B_\mathrm{ext}$) using the Kittel formula\cite{ref:Kittel1951} on both Co/Pt samples, we found a SAW-FMR resonance condition around 4 mT at 1.5 GHz in LiNbO$_{3}$~and 6 mT at 1.3 GHz in Al$_{2}$O$_{3}$.




\begin{thebibliography}{32}%
	\makeatletter
	\providecommand \@ifxundefined [1]{%
		\@ifx{#1\undefined}
	}%
	\providecommand \@ifnum [1]{%
		\ifnum #1\expandafter \@firstoftwo
		\else \expandafter \@secondoftwo
		\fi
	}%
	\providecommand \@ifx [1]{%
		\ifx #1\expandafter \@firstoftwo
		\else \expandafter \@secondoftwo
		\fi
	}%
	\providecommand \natexlab [1]{#1}%
	\providecommand \enquote  [1]{``#1''}%
	\providecommand \bibnamefont  [1]{#1}%
	\providecommand \bibfnamefont [1]{#1}%
	\providecommand \citenamefont [1]{#1}%
	\providecommand \href@noop [0]{\@secondoftwo}%
	\providecommand \href [0]{\begingroup \@sanitize@url \@href}%
	\providecommand \@href[1]{\@@startlink{#1}\@@href}%
	\providecommand \@@href[1]{\endgroup#1\@@endlink}%
	\providecommand \@sanitize@url [0]{\catcode `\\12\catcode `\$12\catcode
		`\&12\catcode `\#12\catcode `\^12\catcode `\_12\catcode `\%12\relax}%
	\providecommand \@@startlink[1]{}%
	\providecommand \@@endlink[0]{}%
	\providecommand \url  [0]{\begingroup\@sanitize@url \@url }%
	\providecommand \@url [1]{\endgroup\@href {#1}{\urlprefix }}%
	\providecommand \urlprefix  [0]{URL }%
	\providecommand \Eprint [0]{\href }%
	\providecommand \doibase [0]{http://dx.doi.org/}%
	\providecommand \selectlanguage [0]{\@gobble}%
	\providecommand \bibinfo  [0]{\@secondoftwo}%
	\providecommand \bibfield  [0]{\@secondoftwo}%
	\providecommand \translation [1]{[#1]}%
	\providecommand \BibitemOpen [0]{}%
	\providecommand \bibitemStop [0]{}%
	\providecommand \bibitemNoStop [0]{.\EOS\space}%
	\providecommand \EOS [0]{\spacefactor3000\relax}%
	\providecommand \BibitemShut  [1]{\csname bibitem#1\endcsname}%
	\let\auto@bib@innerbib\@empty
	\bibitem [{\citenamefont {Baibich}\ \emph {et~al.}(1988)\citenamefont
		{Baibich}, \citenamefont {Broto}, \citenamefont {Fert}, \citenamefont {{Van
				Dau}}, \citenamefont {Petroff}, \citenamefont {Eitenne}, \citenamefont
		{Creuzet}, \citenamefont {Friederich},\ and\ \citenamefont
		{Chazelas}}]{ref:Baibich1988}%
	\BibitemOpen
	\bibfield  {author} {\bibinfo {author} {\bibfnamefont {M.~N.}\ \bibnamefont
			{Baibich}}, \bibinfo {author} {\bibfnamefont {J.~M.}\ \bibnamefont {Broto}},
		\bibinfo {author} {\bibfnamefont {A.}~\bibnamefont {Fert}}, \bibinfo {author}
		{\bibfnamefont {F.~N.}\ \bibnamefont {{Van Dau}}}, \bibinfo {author}
		{\bibfnamefont {F.}~\bibnamefont {Petroff}}, \bibinfo {author} {\bibfnamefont
			{P.}~\bibnamefont {Eitenne}}, \bibinfo {author} {\bibfnamefont
			{G.}~\bibnamefont {Creuzet}}, \bibinfo {author} {\bibfnamefont
			{A.}~\bibnamefont {Friederich}}, \ and\ \bibinfo {author} {\bibfnamefont
			{J.}~\bibnamefont {Chazelas}},\ }\href {\doibase 10.1103/PhysRevLett.61.2472}
	{\bibfield  {journal} {\bibinfo  {journal} {Phys. Rev. Lett.}\ }\textbf
		{\bibinfo {volume} {61}},\ \bibinfo {pages} {2472} (\bibinfo {year}
		{1988})}\BibitemShut {NoStop}%
	\bibitem [{\citenamefont {Binasch}\ \emph {et~al.}(1989)\citenamefont
		{Binasch}, \citenamefont {Gr{\"{u}}nberg}, \citenamefont {Saurenbach},\ and\
		\citenamefont {Zinn}}]{ref:Binasch1989}%
	\BibitemOpen
	\bibfield  {author} {\bibinfo {author} {\bibfnamefont {G.}~\bibnamefont
			{Binasch}}, \bibinfo {author} {\bibfnamefont {P.}~\bibnamefont
			{Gr{\"{u}}nberg}}, \bibinfo {author} {\bibfnamefont {F.}~\bibnamefont
			{Saurenbach}}, \ and\ \bibinfo {author} {\bibfnamefont {W.}~\bibnamefont
			{Zinn}},\ }\href {\doibase 10.1103/PhysRevB.39.4828} {\bibfield  {journal}
		{\bibinfo  {journal} {Phys. Rev. B}\ }\textbf {\bibinfo {volume} {39}},\
		\bibinfo {pages} {4828} (\bibinfo {year} {1989})}\BibitemShut {NoStop}%
	\bibitem [{\citenamefont {George}\ \emph {et~al.}(1994)\citenamefont {George},
		\citenamefont {Pereira}, \citenamefont {Barth{\'{e}}l{\'{e}}my},
		\citenamefont {Petroff}, \citenamefont {Steren}, \citenamefont {Duvail},
		\citenamefont {Fert}, \citenamefont {Loloee}, \citenamefont {Holody},\ and\
		\citenamefont {Schroeder}}]{ref:George1994}%
	\BibitemOpen
	\bibfield  {author} {\bibinfo {author} {\bibfnamefont {J.~M.}\ \bibnamefont
			{George}}, \bibinfo {author} {\bibfnamefont {L.~G.}\ \bibnamefont {Pereira}},
		\bibinfo {author} {\bibfnamefont {A.}~\bibnamefont {Barth{\'{e}}l{\'{e}}my}},
		\bibinfo {author} {\bibfnamefont {F.}~\bibnamefont {Petroff}}, \bibinfo
		{author} {\bibfnamefont {L.}~\bibnamefont {Steren}}, \bibinfo {author}
		{\bibfnamefont {J.~L.}\ \bibnamefont {Duvail}}, \bibinfo {author}
		{\bibfnamefont {A.}~\bibnamefont {Fert}}, \bibinfo {author} {\bibfnamefont
			{R.}~\bibnamefont {Loloee}}, \bibinfo {author} {\bibfnamefont
			{P.}~\bibnamefont {Holody}}, \ and\ \bibinfo {author} {\bibfnamefont {P.~A.}\
			\bibnamefont {Schroeder}},\ }\href {\doibase 10.1103/PhysRevLett.72.408}
	{\bibfield  {journal} {\bibinfo  {journal} {Phys. Rev. Lett.}\ }\textbf
		{\bibinfo {volume} {72}},\ \bibinfo {pages} {408} (\bibinfo {year}
		{1994})}\BibitemShut {NoStop}%
	\bibitem [{\citenamefont {Maekawa}\ \emph {et~al.}(2017)\citenamefont
		{Maekawa}, \citenamefont {Valenzuela}, \citenamefont {Saitoh},\ and\
		\citenamefont {Kimura}}]{ref:Maekawa2017}%
	\BibitemOpen
	\bibfield  {author} {\bibinfo {author} {\bibfnamefont {S.}~\bibnamefont
			{Maekawa}}, \bibinfo {author} {\bibfnamefont {S.~O.}\ \bibnamefont
			{Valenzuela}}, \bibinfo {author} {\bibfnamefont {E.}~\bibnamefont {Saitoh}},
		\ and\ \bibinfo {author} {\bibfnamefont {T.}~\bibnamefont {Kimura}},\
	}\href@noop {} {\bibfield  {journal} {\bibinfo  {journal} {Spin Current
				(Oxford University Press)}\ } (\bibinfo {year} {2017})}\BibitemShut {NoStop}%
	\bibitem [{\citenamefont {Berger}(1996)}]{ref:Berger1996}%
	\BibitemOpen
	\bibfield  {author} {\bibinfo {author} {\bibfnamefont {L.}~\bibnamefont
			{Berger}},\ }\href {\doibase 10.1103/PhysRevB.54.9353} {\bibfield  {journal}
		{\bibinfo  {journal} {Phys. Rev. B}\ }\textbf {\bibinfo {volume} {54}},\
		\bibinfo {pages} {9353} (\bibinfo {year} {1996})}\BibitemShut {NoStop}%
	\bibitem [{\citenamefont {Makarov}\ \emph {et~al.}(2013)\citenamefont
		{Makarov}, \citenamefont {Windbacher}, \citenamefont {Sverdlov},
		\citenamefont {al}, \citenamefont {Wang}, \citenamefont {Alzate},\ and\
		\citenamefont {{Khalili Amiri}}}]{ref:Makarov2013}%
	\BibitemOpen
	\bibfield  {author} {\bibinfo {author} {\bibfnamefont {A.}~\bibnamefont
			{Makarov}}, \bibinfo {author} {\bibfnamefont {T.}~\bibnamefont {Windbacher}},
		\bibinfo {author} {\bibfnamefont {V.}~\bibnamefont {Sverdlov}}, \bibinfo
		{author} {\bibnamefont {al}}, \bibinfo {author} {\bibfnamefont {K.~L.}\
			\bibnamefont {Wang}}, \bibinfo {author} {\bibfnamefont {J.~G.}\ \bibnamefont
			{Alzate}}, \ and\ \bibinfo {author} {\bibfnamefont {P.}~\bibnamefont
			{{Khalili Amiri}}},\ }\href {\doibase 10.1088/0022-3727/46/7/074003}
	{\bibfield  {journal} {\bibinfo  {journal} {J. Phys. D: Appl. Phys}\ }\textbf
		{\bibinfo {volume} {46}},\ \bibinfo {pages} {74003} (\bibinfo {year}
		{2013})}\BibitemShut {NoStop}%
	\bibitem [{\citenamefont {Uchida}\ \emph {et~al.}(2008)\citenamefont {Uchida},
		\citenamefont {Takahashi}, \citenamefont {Harii}, \citenamefont {Ieda},
		\citenamefont {Koshibae}, \citenamefont {Ando}, \citenamefont {Maekawa},\
		and\ \citenamefont {Saitoh}}]{ref:Uchida2008}%
	\BibitemOpen
	\bibfield  {author} {\bibinfo {author} {\bibfnamefont {K.}~\bibnamefont
			{Uchida}}, \bibinfo {author} {\bibfnamefont {S.}~\bibnamefont {Takahashi}},
		\bibinfo {author} {\bibfnamefont {K.}~\bibnamefont {Harii}}, \bibinfo
		{author} {\bibfnamefont {J.}~\bibnamefont {Ieda}}, \bibinfo {author}
		{\bibfnamefont {W.}~\bibnamefont {Koshibae}}, \bibinfo {author}
		{\bibfnamefont {K.}~\bibnamefont {Ando}}, \bibinfo {author} {\bibfnamefont
			{S.}~\bibnamefont {Maekawa}}, \ and\ \bibinfo {author} {\bibfnamefont
			{E.}~\bibnamefont {Saitoh}},\ }\href {\doibase 10.1038/nature07321}
	{\bibfield  {journal} {\bibinfo  {journal} {Nature}\ }\textbf {\bibinfo
			{volume} {455}},\ \bibinfo {pages} {778} (\bibinfo {year}
		{2008})}\BibitemShut {NoStop}%
	\bibitem [{\citenamefont {Uchida}\ \emph {et~al.}(2010)\citenamefont {Uchida},
		\citenamefont {Xiao}, \citenamefont {Adachi}, \citenamefont {Ohe},
		\citenamefont {Takahashi}, \citenamefont {Ieda}, \citenamefont {Ota},
		\citenamefont {Kajiwara}, \citenamefont {Umezawa}, \citenamefont {Kawai},
		\citenamefont {Bauer}, \citenamefont {Maekawa},\ and\ \citenamefont
		{Saitoh}}]{ref:Uchida2010}%
	\BibitemOpen
	\bibfield  {author} {\bibinfo {author} {\bibfnamefont {K.}~\bibnamefont
			{Uchida}}, \bibinfo {author} {\bibfnamefont {J.}~\bibnamefont {Xiao}},
		\bibinfo {author} {\bibfnamefont {H.}~\bibnamefont {Adachi}}, \bibinfo
		{author} {\bibfnamefont {J.}~\bibnamefont {Ohe}}, \bibinfo {author}
		{\bibfnamefont {S.}~\bibnamefont {Takahashi}}, \bibinfo {author}
		{\bibfnamefont {J.}~\bibnamefont {Ieda}}, \bibinfo {author} {\bibfnamefont
			{T.}~\bibnamefont {Ota}}, \bibinfo {author} {\bibfnamefont {Y.}~\bibnamefont
			{Kajiwara}}, \bibinfo {author} {\bibfnamefont {H.}~\bibnamefont {Umezawa}},
		\bibinfo {author} {\bibfnamefont {H.}~\bibnamefont {Kawai}}, \bibinfo
		{author} {\bibfnamefont {G.~E.}\ \bibnamefont {Bauer}}, \bibinfo {author}
		{\bibfnamefont {S.}~\bibnamefont {Maekawa}}, \ and\ \bibinfo {author}
		{\bibfnamefont {E.}~\bibnamefont {Saitoh}},\ }\href {\doibase
		10.1038/nmat2856} {\bibfield  {journal} {\bibinfo  {journal} {Nature
				Materials}\ }\textbf {\bibinfo {volume} {9}},\ \bibinfo {pages} {894}
		(\bibinfo {year} {2010})}\BibitemShut {NoStop}%
	\bibitem [{\citenamefont {Matsuo}\ \emph {et~al.}(2018)\citenamefont {Matsuo},
		\citenamefont {Ohnuma}, \citenamefont {Kato},\ and\ \citenamefont
		{Maekawa}}]{ref:Matsuo2018}%
	\BibitemOpen
	\bibfield  {author} {\bibinfo {author} {\bibfnamefont {M.}~\bibnamefont
			{Matsuo}}, \bibinfo {author} {\bibfnamefont {Y.}~\bibnamefont {Ohnuma}},
		\bibinfo {author} {\bibfnamefont {T.}~\bibnamefont {Kato}}, \ and\ \bibinfo
		{author} {\bibfnamefont {S.}~\bibnamefont {Maekawa}},\ }\href {\doibase
		10.1103/PhysRevLett.120.037201} {\bibfield  {journal} {\bibinfo  {journal}
			{Phys. Rev. Lett.}\ }\textbf {\bibinfo {volume} {120}},\ \bibinfo {pages}
		{037201} (\bibinfo {year} {2018})}\BibitemShut {NoStop}%
	\bibitem [{\citenamefont {Kimura}\ \emph {et~al.}(2007)\citenamefont {Kimura},
		\citenamefont {Otani}, \citenamefont {Sato}, \citenamefont {Takahashi},\ and\
		\citenamefont {Maekawa}}]{ref:Kimura2007}%
	\BibitemOpen
	\bibfield  {author} {\bibinfo {author} {\bibfnamefont {T.}~\bibnamefont
			{Kimura}}, \bibinfo {author} {\bibfnamefont {Y.}~\bibnamefont {Otani}},
		\bibinfo {author} {\bibfnamefont {T.}~\bibnamefont {Sato}}, \bibinfo {author}
		{\bibfnamefont {S.}~\bibnamefont {Takahashi}}, \ and\ \bibinfo {author}
		{\bibfnamefont {S.}~\bibnamefont {Maekawa}},\ }\href {\doibase
		10.1103/PhysRevLett.98.156601} {\bibfield  {journal} {\bibinfo  {journal}
			{Phys. Rev. Lett.}\ }\textbf {\bibinfo {volume} {98}},\ \bibinfo {pages}
		{156601} (\bibinfo {year} {2007})}\BibitemShut {NoStop}%
	\bibitem [{\citenamefont {Weiler}\ \emph {et~al.}(2012)\citenamefont {Weiler},
		\citenamefont {Huebl}, \citenamefont {Goerg}, \citenamefont {Czeschka},
		\citenamefont {Gross},\ and\ \citenamefont {Goennenwein}}]{ref:Weiler2012}%
	\BibitemOpen
	\bibfield  {author} {\bibinfo {author} {\bibfnamefont {M.}~\bibnamefont
			{Weiler}}, \bibinfo {author} {\bibfnamefont {H.}~\bibnamefont {Huebl}},
		\bibinfo {author} {\bibfnamefont {F.~S.}\ \bibnamefont {Goerg}}, \bibinfo
		{author} {\bibfnamefont {F.~D.}\ \bibnamefont {Czeschka}}, \bibinfo {author}
		{\bibfnamefont {R.}~\bibnamefont {Gross}}, \ and\ \bibinfo {author}
		{\bibfnamefont {S.~T.~B.}\ \bibnamefont {Goennenwein}},\ }\href {\doibase
		10.1103/PhysRevLett.108.176601} {\bibfield  {journal} {\bibinfo  {journal}
			{Phys. Rev. Lett.}\ }\textbf {\bibinfo {volume} {108}},\ \bibinfo {pages}
		{176601} (\bibinfo {year} {2012})}\BibitemShut {NoStop}%
	\bibitem [{\citenamefont {Ando}\ \emph {et~al.}(2011)\citenamefont {Ando},
		\citenamefont {Takahashi}, \citenamefont {Ieda}, \citenamefont {Kajiwara},
		\citenamefont {Nakayama}, \citenamefont {Yoshino}, \citenamefont {Harii},
		\citenamefont {Fujikawa}, \citenamefont {Matsuo}, \citenamefont {Maekawa},\
		and\ \citenamefont {Saitoh}}]{ref:Ando2011}%
	\BibitemOpen
	\bibfield  {author} {\bibinfo {author} {\bibfnamefont {K.}~\bibnamefont
			{Ando}}, \bibinfo {author} {\bibfnamefont {S.}~\bibnamefont {Takahashi}},
		\bibinfo {author} {\bibfnamefont {J.}~\bibnamefont {Ieda}}, \bibinfo {author}
		{\bibfnamefont {Y.}~\bibnamefont {Kajiwara}}, \bibinfo {author}
		{\bibfnamefont {H.}~\bibnamefont {Nakayama}}, \bibinfo {author}
		{\bibfnamefont {T.}~\bibnamefont {Yoshino}}, \bibinfo {author} {\bibfnamefont
			{K.}~\bibnamefont {Harii}}, \bibinfo {author} {\bibfnamefont
			{Y.}~\bibnamefont {Fujikawa}}, \bibinfo {author} {\bibfnamefont
			{M.}~\bibnamefont {Matsuo}}, \bibinfo {author} {\bibfnamefont
			{S.}~\bibnamefont {Maekawa}}, \ and\ \bibinfo {author} {\bibfnamefont
			{E.}~\bibnamefont {Saitoh}},\ }\href {https://doi.org/10.1063/1.3587173}
	{\bibfield  {journal} {\bibinfo  {journal} {J. Appl. Phys}\ }\textbf
		{\bibinfo {volume} {109}},\ \bibinfo {pages} {103913} (\bibinfo {year}
		{2011})}\BibitemShut {NoStop}%
	\bibitem [{\citenamefont {Okada}\ \emph {et~al.}(2019)\citenamefont {Okada},
		\citenamefont {Takeuchi}, \citenamefont {Furuya}, \citenamefont {Zhang},
		\citenamefont {Sato}, \citenamefont {Fukami},\ and\ \citenamefont
		{Ohno}}]{ref:Okada2019}%
	\BibitemOpen
	\bibfield  {author} {\bibinfo {author} {\bibfnamefont {A.}~\bibnamefont
			{Okada}}, \bibinfo {author} {\bibfnamefont {Y.}~\bibnamefont {Takeuchi}},
		\bibinfo {author} {\bibfnamefont {K.}~\bibnamefont {Furuya}}, \bibinfo
		{author} {\bibfnamefont {C.}~\bibnamefont {Zhang}}, \bibinfo {author}
		{\bibfnamefont {H.}~\bibnamefont {Sato}}, \bibinfo {author} {\bibfnamefont
			{S.}~\bibnamefont {Fukami}}, \ and\ \bibinfo {author} {\bibfnamefont
			{H.}~\bibnamefont {Ohno}},\ }\href@noop {} {\bibfield  {journal} {\bibinfo
			{journal} {PR Applied}\ }\textbf {\bibinfo {volume} {12}},\ \bibinfo {pages}
		{014040} (\bibinfo {year} {2019})}\BibitemShut {NoStop}%
	\bibitem [{\citenamefont {Weiler}\ \emph {et~al.}(2011)\citenamefont {Weiler},
		\citenamefont {Dreher}, \citenamefont {Heeg}, \citenamefont {Huebl},
		\citenamefont {Gross}, \citenamefont {Brandt},\ and\ \citenamefont
		{Goennenwein}}]{ref:Weiler2011}%
	\BibitemOpen
	\bibfield  {author} {\bibinfo {author} {\bibfnamefont {M.}~\bibnamefont
			{Weiler}}, \bibinfo {author} {\bibfnamefont {L.}~\bibnamefont {Dreher}},
		\bibinfo {author} {\bibfnamefont {C.}~\bibnamefont {Heeg}}, \bibinfo {author}
		{\bibfnamefont {H.}~\bibnamefont {Huebl}}, \bibinfo {author} {\bibfnamefont
			{R.}~\bibnamefont {Gross}}, \bibinfo {author} {\bibfnamefont {M.~S.}\
			\bibnamefont {Brandt}}, \ and\ \bibinfo {author} {\bibfnamefont {S.~T.}\
			\bibnamefont {Goennenwein}},\ }\href@noop {} {\bibfield  {journal} {\bibinfo
			{journal} {Phys. Rev. Lett.}\ }\textbf {\bibinfo {volume} {106}},\ \bibinfo
		{pages} {117601} (\bibinfo {year} {2011})}\BibitemShut {NoStop}%
	\bibitem [{\citenamefont {Thevenard}\ \emph {et~al.}(2016)\citenamefont
		{Thevenard}, \citenamefont {Camara}, \citenamefont {Majrab}, \citenamefont
		{Bernard}, \citenamefont {Rovillain}, \citenamefont {Lema{\^{i}}tre},
		\citenamefont {Gourdon},\ and\ \citenamefont {Duquesne}}]{ref:Thevenard2016}%
	\BibitemOpen
	\bibfield  {author} {\bibinfo {author} {\bibfnamefont {L.}~\bibnamefont
			{Thevenard}}, \bibinfo {author} {\bibfnamefont {I.~S.}\ \bibnamefont
			{Camara}}, \bibinfo {author} {\bibfnamefont {S.}~\bibnamefont {Majrab}},
		\bibinfo {author} {\bibfnamefont {M.}~\bibnamefont {Bernard}}, \bibinfo
		{author} {\bibfnamefont {P.}~\bibnamefont {Rovillain}}, \bibinfo {author}
		{\bibfnamefont {A.}~\bibnamefont {Lema{\^{i}}tre}}, \bibinfo {author}
		{\bibfnamefont {C.}~\bibnamefont {Gourdon}}, \ and\ \bibinfo {author}
		{\bibfnamefont {J.-Y.}\ \bibnamefont {Duquesne}},\ }\href@noop {} {\bibfield
		{journal} {\bibinfo  {journal} {Phys. Rev. B}\ }\textbf {\bibinfo {volume}
			{93}} (\bibinfo {year} {2016})}\BibitemShut {NoStop}%
	\bibitem [{\citenamefont {Duquesne}\ \emph {et~al.}(2019)\citenamefont
		{Duquesne}, \citenamefont {Rovillain}, \citenamefont {Hepburn}, \citenamefont
		{Eddrief}, \citenamefont {Atkinson}, \citenamefont {Anane}, \citenamefont
		{Ranchal},\ and\ \citenamefont {Marangolo}}]{ref:Duquesne2019}%
	\BibitemOpen
	\bibfield  {author} {\bibinfo {author} {\bibfnamefont {J.-Y.}\ \bibnamefont
			{Duquesne}}, \bibinfo {author} {\bibfnamefont {P.}~\bibnamefont {Rovillain}},
		\bibinfo {author} {\bibfnamefont {C.}~\bibnamefont {Hepburn}}, \bibinfo
		{author} {\bibfnamefont {M.}~\bibnamefont {Eddrief}}, \bibinfo {author}
		{\bibfnamefont {P.}~\bibnamefont {Atkinson}}, \bibinfo {author}
		{\bibfnamefont {A.}~\bibnamefont {Anane}}, \bibinfo {author} {\bibfnamefont
			{R.}~\bibnamefont {Ranchal}}, \ and\ \bibinfo {author} {\bibfnamefont
			{M.}~\bibnamefont {Marangolo}},\ }\href {\doibase
		10.1103/PhysRevApplied.12.024042} {\bibfield  {journal} {\bibinfo  {journal}
			{PR Applied}\ }\textbf {\bibinfo {volume} {12}},\ \bibinfo {pages} {024042}
		(\bibinfo {year} {2019})}\BibitemShut {NoStop}%
	\bibitem [{\citenamefont {Adachi}\ and\ \citenamefont
		{Maekawa}(2014)}]{ref:Adachi2014}%
	\BibitemOpen
	\bibfield  {author} {\bibinfo {author} {\bibfnamefont {H.}~\bibnamefont
			{Adachi}}\ and\ \bibinfo {author} {\bibfnamefont {S.}~\bibnamefont
			{Maekawa}},\ }\href@noop {} {\bibfield  {journal} {\bibinfo  {journal} {Solid
				State Communications}\ }\textbf {\bibinfo {volume} {198}},\ \bibinfo {pages}
		{22} (\bibinfo {year} {2014})}\BibitemShut {NoStop}%
	\bibitem [{\citenamefont {Saitoh}\ \emph {et~al.}(2006)\citenamefont {Saitoh},
		\citenamefont {Ueda}, \citenamefont {Miyajima},\ and\ \citenamefont
		{Tatara}}]{ref:Saitoh2006}%
	\BibitemOpen
	\bibfield  {author} {\bibinfo {author} {\bibfnamefont {E.}~\bibnamefont
			{Saitoh}}, \bibinfo {author} {\bibfnamefont {M.}~\bibnamefont {Ueda}},
		\bibinfo {author} {\bibfnamefont {H.}~\bibnamefont {Miyajima}}, \ and\
		\bibinfo {author} {\bibfnamefont {G.}~\bibnamefont {Tatara}},\ }\href
	{\doibase 10.1063/1.2199473} {\bibfield  {journal} {\bibinfo  {journal}
			{Appl. Phys. Lett.}\ }\textbf {\bibinfo {volume} {88}},\ \bibinfo {pages}
		{182509} (\bibinfo {year} {2006})}\BibitemShut {NoStop}%
	\bibitem [{\citenamefont {Nassau}\ \emph
		{et~al.}(1966{\natexlab{a}})\citenamefont {Nassau}, \citenamefont
		{Levinstein},\ and\ \citenamefont {Loiacono}}]{ref:Nassau1966a}%
	\BibitemOpen
	\bibfield  {author} {\bibinfo {author} {\bibfnamefont {K.}~\bibnamefont
			{Nassau}}, \bibinfo {author} {\bibfnamefont {H.~J.}\ \bibnamefont
			{Levinstein}}, \ and\ \bibinfo {author} {\bibfnamefont {G.~M.}\ \bibnamefont
			{Loiacono}},\ }\href {\doibase 10.1016/0022-3697(66)90070-9} {\bibfield
		{journal} {\bibinfo  {journal} {Journal of Physics and Chemistry of Solids}\
		}\textbf {\bibinfo {volume} {27}},\ \bibinfo {pages} {983} (\bibinfo {year}
		{1966}{\natexlab{a}})}\BibitemShut {NoStop}%
	\bibitem [{\citenamefont {Nassau}\ \emph
		{et~al.}(1966{\natexlab{b}})\citenamefont {Nassau}, \citenamefont
		{Levinstein},\ and\ \citenamefont {Loiacono}}]{ref:Nassau1966b}%
	\BibitemOpen
	\bibfield  {author} {\bibinfo {author} {\bibfnamefont {K.}~\bibnamefont
			{Nassau}}, \bibinfo {author} {\bibfnamefont {H.~J.}\ \bibnamefont
			{Levinstein}}, \ and\ \bibinfo {author} {\bibfnamefont {G.~M.}\ \bibnamefont
			{Loiacono}},\ }\href {\doibase 10.1016/0022-3697(66)90071-0} {\bibfield
		{journal} {\bibinfo  {journal} {Journal of Physics and Chemistry of Solids}\
		}\textbf {\bibinfo {volume} {27}},\ \bibinfo {pages} {989} (\bibinfo {year}
		{1966}{\natexlab{b}})}\BibitemShut {NoStop}%
	\bibitem [{\citenamefont {Hamida}\ \emph {et~al.}(2013)\citenamefont {Hamida},
		\citenamefont {Sievers}, \citenamefont {Pierz},\ and\ \citenamefont
		{Schumacher}}]{ref:Hamida2013}%
	\BibitemOpen
	\bibfield  {author} {\bibinfo {author} {\bibfnamefont {A.~B.}\ \bibnamefont
			{Hamida}}, \bibinfo {author} {\bibfnamefont {S.}~\bibnamefont {Sievers}},
		\bibinfo {author} {\bibfnamefont {K.}~\bibnamefont {Pierz}}, \ and\ \bibinfo
		{author} {\bibfnamefont {H.~W.}\ \bibnamefont {Schumacher}},\ }\href
	{\doibase 10.1063/1.4823740} {\bibfield  {journal} {\bibinfo  {journal} {J.
				Appl. Phys.}\ }\textbf {\bibinfo {volume} {114}},\ \bibinfo {pages} {123704}
		(\bibinfo {year} {2013})}\BibitemShut {NoStop}%
	\bibitem [{\citenamefont {Kittel}(1951)}]{ref:Kittel1951}%
	\BibitemOpen
	\bibfield  {author} {\bibinfo {author} {\bibfnamefont {C.}~\bibnamefont
			{Kittel}},\ }\href {http://dx.doi.org/10.1051/jphysrad:01951001203029100}
	{\bibfield  {journal} {\bibinfo  {journal} {J. Phys. Radium}\ }\textbf
		{\bibinfo {volume} {12}},\ \bibinfo {pages} {291} (\bibinfo {year}
		{1951})}\BibitemShut {NoStop}%
	\bibitem [{\citenamefont {Czeschka}\ \emph {et~al.}(2011)\citenamefont
		{Czeschka}, \citenamefont {Dreher}, \citenamefont {Brandt}, \citenamefont
		{Weiler}, \citenamefont {Althammer}, \citenamefont {Imort}, \citenamefont
		{Reiss}, \citenamefont {Thomas}, \citenamefont {Schoch}, \citenamefont
		{Limmer}, \citenamefont {Huebl}, \citenamefont {Gross},\ and\ \citenamefont
		{Goennenwein}}]{ref:Czeschka2011}%
	\BibitemOpen
	\bibfield  {author} {\bibinfo {author} {\bibfnamefont {F.~D.}\ \bibnamefont
			{Czeschka}}, \bibinfo {author} {\bibfnamefont {L.}~\bibnamefont {Dreher}},
		\bibinfo {author} {\bibfnamefont {M.~S.}\ \bibnamefont {Brandt}}, \bibinfo
		{author} {\bibfnamefont {M.}~\bibnamefont {Weiler}}, \bibinfo {author}
		{\bibfnamefont {M.}~\bibnamefont {Althammer}}, \bibinfo {author}
		{\bibfnamefont {I.-M.}\ \bibnamefont {Imort}}, \bibinfo {author}
		{\bibfnamefont {G.}~\bibnamefont {Reiss}}, \bibinfo {author} {\bibfnamefont
			{A.}~\bibnamefont {Thomas}}, \bibinfo {author} {\bibfnamefont
			{W.}~\bibnamefont {Schoch}}, \bibinfo {author} {\bibfnamefont
			{W.}~\bibnamefont {Limmer}}, \bibinfo {author} {\bibfnamefont
			{H.}~\bibnamefont {Huebl}}, \bibinfo {author} {\bibfnamefont
			{R.}~\bibnamefont {Gross}}, \ and\ \bibinfo {author} {\bibfnamefont
			{S.~T.~B.}\ \bibnamefont {Goennenwein}},\ }\href {\doibase
		10.1103/PhysRevLett.107.046601} {\bibfield  {journal} {\bibinfo  {journal}
			{Phys. Rev. Lett.}\ }\textbf {\bibinfo {volume} {107}},\ \bibinfo {pages}
		{046601} (\bibinfo {year} {2011})}\BibitemShut {NoStop}%
	\bibitem [{ref()}]{ref:footnote1}%
	\BibitemOpen
	\href@noop {} {}\bibinfo {note} {The power dependence measurement is
		performed for $\phi = 50^{\circ}$ to get high $V_I$ intensity and the
		possibility to recover the signal at low power. The current emitted at B$_c$
		is also higher but does not disturb the measurement.}\BibitemShut {Stop}%
	\bibitem [{\citenamefont {Camley}(1987)}]{ref:Camley1987}%
	\BibitemOpen
	\bibfield  {author} {\bibinfo {author} {\bibfnamefont {R.~E.}\ \bibnamefont
			{Camley}},\ }\href {\doibase 10.1016/0167-5729(87)90006-9} {\bibfield
		{journal} {\bibinfo  {journal} {Surface Science Reports}\ }\textbf {\bibinfo
			{volume} {7}},\ \bibinfo {pages} {103} (\bibinfo {year} {1987})}\BibitemShut
	{NoStop}%
	\bibitem [{\citenamefont {Velev}\ \emph {et~al.}(2011)\citenamefont {Velev},
		\citenamefont {Jaswal},\ and\ \citenamefont {Tsymbal}}]{ref:Velev2011}%
	\BibitemOpen
	\bibfield  {author} {\bibinfo {author} {\bibfnamefont {J.~P.}\ \bibnamefont
			{Velev}}, \bibinfo {author} {\bibfnamefont {S.~S.}\ \bibnamefont {Jaswal}}, \
		and\ \bibinfo {author} {\bibfnamefont {E.~Y.}\ \bibnamefont {Tsymbal}},\
	}\href@noop {} {\bibfield  {journal} {\bibinfo  {journal} {Philosophical
				Transactions of the Royal Society A: Mathematical, Physical and Engineering
				Sciences}\ }\textbf {\bibinfo {volume} {369}},\ \bibinfo {pages} {3069}
		(\bibinfo {year} {2011})}\BibitemShut {NoStop}%
	\bibitem [{\citenamefont {Jedrecy}\ \emph {et~al.}(2013)\citenamefont
		{Jedrecy}, \citenamefont {von Bardeleben}, \citenamefont {Badjeck},
		\citenamefont {Demaille}, \citenamefont {Stanescu}, \citenamefont {Magnan},\
		and\ \citenamefont {Barbier}}]{ref:Jedrecy2013}%
	\BibitemOpen
	\bibfield  {author} {\bibinfo {author} {\bibfnamefont {N.}~\bibnamefont
			{Jedrecy}}, \bibinfo {author} {\bibfnamefont {H.~J.}\ \bibnamefont {von
				Bardeleben}}, \bibinfo {author} {\bibfnamefont {V.}~\bibnamefont {Badjeck}},
		\bibinfo {author} {\bibfnamefont {D.}~\bibnamefont {Demaille}}, \bibinfo
		{author} {\bibfnamefont {D.}~\bibnamefont {Stanescu}}, \bibinfo {author}
		{\bibfnamefont {H.}~\bibnamefont {Magnan}}, \ and\ \bibinfo {author}
		{\bibfnamefont {A.}~\bibnamefont {Barbier}},\ }\href {\doibase
		10.1103/PhysRevB.88.121409} {\bibfield  {journal} {\bibinfo  {journal} {Phys.
				Rev. B}\ }\textbf {\bibinfo {volume} {88}},\ \bibinfo {pages} {121409(R)}
		(\bibinfo {year} {2013})}\BibitemShut {NoStop}%
	\bibitem [{\citenamefont {Jia}\ \emph {et~al.}(2014)\citenamefont {Jia},
		\citenamefont {Wei}, \citenamefont {Jiang}, \citenamefont {Xue},
		\citenamefont {Sukhov},\ and\ \citenamefont {Berakdar}}]{ref:Jia2014}%
	\BibitemOpen
	\bibfield  {author} {\bibinfo {author} {\bibfnamefont {C.-L.}\ \bibnamefont
			{Jia}}, \bibinfo {author} {\bibfnamefont {T.-L.}\ \bibnamefont {Wei}},
		\bibinfo {author} {\bibfnamefont {C.-J.}\ \bibnamefont {Jiang}}, \bibinfo
		{author} {\bibfnamefont {D.-S.}\ \bibnamefont {Xue}}, \bibinfo {author}
		{\bibfnamefont {A.}~\bibnamefont {Sukhov}}, \ and\ \bibinfo {author}
		{\bibfnamefont {J.}~\bibnamefont {Berakdar}},\ }\href {\doibase
		10.1103/PhysRevB.90.054423} {\bibfield  {journal} {\bibinfo  {journal} {Phys.
				Rev. B}\ }\textbf {\bibinfo {volume} {90}},\ \bibinfo {pages} {054423}
		(\bibinfo {year} {2014})}\BibitemShut {NoStop}%
	\bibitem [{\citenamefont {Chiba}\ \emph {et~al.}(2016)\citenamefont {Chiba},
		\citenamefont {Shibata},\ and\ \citenamefont {Tsukazaki}}]{ref:Chiba2016}%
	\BibitemOpen
	\bibfield  {author} {\bibinfo {author} {\bibfnamefont {D.}~\bibnamefont
			{Chiba}}, \bibinfo {author} {\bibfnamefont {N.}~\bibnamefont {Shibata}}, \
		and\ \bibinfo {author} {\bibfnamefont {A.}~\bibnamefont {Tsukazaki}},\ }\href
	{https://doi.org/10.1038/srep38005} {\bibfield  {journal} {\bibinfo
			{journal} {Sci. Rep.}\ }\textbf {\bibinfo {volume} {6}},\ \bibinfo {pages}
		{38005} (\bibinfo {year} {2016})}\BibitemShut {NoStop}%
	\bibitem [{\citenamefont {Kim}\ \emph {et~al.}(2002)\citenamefont {Kim},
		\citenamefont {Gopalan},\ and\ \citenamefont {Gruverman}}]{ref:Kim2002}%
	\BibitemOpen
	\bibfield  {author} {\bibinfo {author} {\bibfnamefont {S.}~\bibnamefont
			{Kim}}, \bibinfo {author} {\bibfnamefont {V.}~\bibnamefont {Gopalan}}, \ and\
		\bibinfo {author} {\bibfnamefont {A.}~\bibnamefont {Gruverman}},\ }\href
	{https://doi.org/10.1063/1.1470247} {\bibfield  {journal} {\bibinfo
			{journal} {Appl. Phys. Lett.}\ }\textbf {\bibinfo {volume} {80}},\ \bibinfo
		{pages} {2740} (\bibinfo {year} {2002})}\BibitemShut {NoStop}%
	\bibitem [{\citenamefont {Zalessky}\ and\ \citenamefont
		{Fregatov}(2006)}]{ref:Zalessky2006}%
	\BibitemOpen
	\bibfield  {author} {\bibinfo {author} {\bibfnamefont {V.~G.}\ \bibnamefont
			{Zalessky}}\ and\ \bibinfo {author} {\bibfnamefont {S.~O.}\ \bibnamefont
			{Fregatov}},\ }\href {\doibase https://doi.org/10.1016/j.physb.2005.10.097}
	{\bibfield  {journal} {\bibinfo  {journal} {Physica B: Condensed Matter}\
		}\textbf {\bibinfo {volume} {371}},\ \bibinfo {pages} {158 } (\bibinfo {year}
		{2006})}\BibitemShut {NoStop}%
	\bibitem [{\citenamefont {Kiselev}\ \emph {et~al.}(2012)\citenamefont
		{Kiselev}, \citenamefont {Bykov}, \citenamefont {Zhukov}, \citenamefont
		{Antipov}, \citenamefont {Malinkovich},\ and\ \citenamefont
		{Parkhomenko}}]{Kiselev2012}%
	\BibitemOpen
	\bibfield  {author} {\bibinfo {author} {\bibfnamefont {D.~A.}\ \bibnamefont
			{Kiselev}}, \bibinfo {author} {\bibfnamefont {A.~S.}\ \bibnamefont {Bykov}},
		\bibinfo {author} {\bibfnamefont {R.~N.}\ \bibnamefont {Zhukov}}, \bibinfo
		{author} {\bibfnamefont {V.~V.}\ \bibnamefont {Antipov}}, \bibinfo {author}
		{\bibfnamefont {M.~D.}\ \bibnamefont {Malinkovich}}, \ and\ \bibinfo {author}
		{\bibfnamefont {Y.~N.}\ \bibnamefont {Parkhomenko}},\ }\href {\doibase
		10.1134/S1063774512060041} {\bibfield  {journal} {\bibinfo  {journal}
			{Crystallography Reports}\ }\textbf {\bibinfo {volume} {57}},\ \bibinfo
		{pages} {781} (\bibinfo {year} {2012})}\BibitemShut {NoStop}%
\end{thebibliography}


%


\end{document}